%% file: kou08multi_paper.tex
\documentclass{article}
\usepackage{nips07submit_e,times}

\title{ A multi-resolution, non-parametric, Bayesian framework  for  identification of spatially-varying model parameters.
}

\author{
P.S. Koutsourelakis  \\
School of Civil and Environmental Engineering \& Center for Applied Mathematics\\
Cornell University \\
\texttt{pk285@cornell.edu} \\
}

\usepackage{amsmath,amssymb}
\usepackage[dvips]{psfrag,graphicx}
\usepackage{color}
\usepackage{subfigure}
\usepackage{here}

\newcommand{\ee}{\end{equation}}
\newcommand{\be}{\begin{equation}}
\newcommand{\ec}{\end{center}}
\newcommand{\bc}{\begin{center}}
\newcommand{\eea}{\end{eqnarray}}
\newcommand{\bea}{\begin{eqnarray}}
\newcommand{\bd}{\begin{description}}
\newcommand{\ed}{\end{description}}
\newcommand{\bi}{\begin{itemize}}
\newcommand{\ei}{\end{itemize}}

\newcommand{\bs}{\boldsymbol}
\def\RR{ \mathbb R}
\newcommand{\refeq}[1]{Equation (\ref{#1})}
\begin{document}

\makeanontitle

\begin{abstract}
\input{abstract}

\end{abstract}

\section{Introduction}
\input{introduction}

\section{Problem Definition \& Motivation}
\input{definition}

\section{Methodology}
\input{methodology}

\section{Numerical Examples}
\input{results_short}

\section{Conclusions}
\input{conclusions}


\clearpage
\newpage

\bibliographystyle{plain}
\bibliography{nsf_career_proposal}

\end{document}

%% file: abstract.tex
This paper proposes a hierarchical, multi-resolution framework for the identification of model parameters and their spatially variability from noisy measurements of the response or output. Such parameters are frequently encountered in PDE-based  models and correspond to  quantities such as density or pressure fields, elasto-plastic moduli and internal variables in solid mechanics, conductivity fields in heat diffusion problems, permeability fields in fluid flow through porous media etc. The proposed model has all the advantages of traditional Bayesian formulations such as the ability to produce measures of confidence for the inferences made and  providing  not only predictive estimates but also quantitative measures of the  predictive uncertainty. In contrast to existing approaches it utilizes a parsimonious, non-parametric formulation  that favors sparse representations and whose complexity can be determined from the data. The proposed framework in non-intrusive and makes use of a sequence of forward solvers operating at various resolutions. As a result, inexpensive, coarse solvers are used to identify the most salient features of the unknown field(s) which are subsequently enriched by invoking solvers operating at finer resolutions.  This leads to significant computational savings particularly in problems involving computationally demanding forward models but also improvements in accuracy. It is based on a novel, adaptive scheme based on Sequential Monte Carlo sampling  which is embarrassingly parallelizable and circumvents issues with slow mixing encountered in Markov Chain Monte Carlo schemes. The capabilities of the proposed methodology  are illustrated in problems from nonlinear solid mechanics with special attention to cases where the data is contaminated with random noise and the scale of variability of the unknown field is smaller than the scale of the grid where observations are collected.

%% file: introduction.tex
The prodigious advances in computational modeling of physical processes and the development of highly non-linear, multiscale and multiphysics models poses several challenges in parameter identification.  We are frequently using large, forward  models which imply a significant computational burden, in order to analyze complex phenomena.The extensive use of such models  poses several challenges in parameter identification as the accuracy of the results provided depends strongly on assigning proper values to the various model parameters. In mechanics of materials, accurate mechanical property identification can guide damage  detection  and an informed assessment of the system's reliability (\cite{joh05mul}). Identifying property-cross correlations can lead to the design of multi-functional materials (\cite{tor02ran}).
In biomechanics, the detection of variations in mechanical properties of human tissue can reveal the appearance of diseases (arteriosclerosis, malignant tumors) but  can also  be used to assess the effectivity of various treatments (\cite{bri07inv,Fatemi:1998}). Permeability estimation for soil transport processes can assist in detection of contaminants, oil exploration etc. (\cite{Wang:2006,Fienen:2004}). 


We consider phenomena described by a set of (coupled) elliptic, parabolic or hyperbolic PDEs and associated boundary (and initial) conditions:
\be
\label{eq:0}
\mathcal{A} (\bs{y}(\bs{x}) ; f(\bs{x}) ) = 0, \qquad \forall x \in \mathcal{D}
\ee
 where $\mathcal{A}$ denotes the differential  operator defined on a domain  $\mathcal{D} \in \RR^d$ , where $d$ is the number of spatial dimensions. $\mathcal{A}$  depends on spatially varying  coefficients $f(\bs{x})$, $\bs{x} \in \mathcal{D}$. Advances in computational mathematics have given rise to several efficient solvers for a wide-range of such systems and have revolutionized simulation-based analysis and design (\cite{nsf2006}). Our primary interest is to identify $f(\bs{x})$ from a set of (potentially noisy) measurements of the response $\bs{y}_i=\bs{y}(\bs{x}_i)$ at a number of distinct locations $\bs{x}_i \in \mathcal{D}$. In the case of time-dependent PDEs, the available data might also be indexed by time. 
 Several different processes in solid and fluid mechanics, transport phenomena, heat diffusion etc fall under this general setting and even though the coefficients $f(\bs{x})$ have a different physical interpretation, the associated inverse  problems  exhibit similar mathematical characteristics.


Two basic approaches have been followed in addressing problems of data-driven  parametric identification. On one  hand, deterministic optimization techniques which attempt to minimize the sum of the  squares of the deviations between model predictions and observations. Gradient or global, intrusive or non-intrusive  techniques are introduced for performing the optimization task.   Usually the objective function is augmented with regularization terms (e.g. Tikhonov regularization \cite{tik63sol}) which alleviate issues with the ill-posednesss of the problem (\cite{tik77sol,gro93inv,eng96reg,vog02com,Calvetti:2003,kai05com}). Such  deterministic inverse techniques based on exact matching or least-squares optimization, lead to point estimates of unknowns without rigorously considering the statistical nature of system uncertainties and  without providing quantification of the uncertainty in the inverse solution.

 The direct stochastic counterpart of optimization methods involves frequentist approaches  based on maximum likelihood estimators that aim at  maximizing the probability of observations given the inverse solution maximum (\cite{fad95unc,eme00unc}).  In recent years significant attention has been directed towards statistical approaches based on the Bayesian paradigm which attempt to calculate a (posterior) probability distribution function on the parameters of interest. Bayesian formulations  offer several advantages as they  provide a unified framework for dealing with the uncertainty introduced by the incomplete and noisy measurements and assessing quantitatively resulting inferential uncertainties. Significant successes have been noted in applications  such as medical tomography (\cite{Weir:1997}), geological tomography (\cite{gla04sto,and01bay}), hydrology (\cite{Lee:2002a}), petroleum engineering (\cite{Hegstad:2001,Craig:2001}), as well as a host of other physical, biological, or social systems (\cite{Kitanidis:1986,Schmidt:1999,Wang:2005a,Liu:2008}).

  Identification of spatially varying model parameters  poses several modeling  and computational issues.  Representations of the parametric fields in existing approaches artificially impose a minimum length scale of variability usually determined by the discretization size of the governing PDEs (\cite{Lee:2002a}). Furthermore, they are associated with a very large  vector of unknowns. Inference in high-dimensional spaces using standard optimization or sampling schemes (e.g. Markov Chain Monte Carlo (MCMC)),  is generally impractical as it requires an  exuberant number of calls to the forward simulator in order to achieve convergence. Particularly in  Bayesian formulations where the inference results are  much richer and involve a distribution rather than a single value for the parameters of interest, the computational effort implied by repeated calls to the forward solver can be enormous and constitute the method impractical for realistic applications. These problems are amplified if the posterior distribution is multi-modal i.e. several significantly different scenaria are likely given the available data.  While it is apparent that, computationally inexpensive,  coarser scale simulations can assist the identification process (\cite{Dostert:2006}), the critical task of efficiently transferring the information across resolutions still remains (\cite{liu00gen,hig03mar,Wang:2006}). Previous attempts using  parallel tempering (e.g. \cite{hol06mul})  or hierarchical representations based on Markov trees (\cite{wan06per})  require performing inference on representations at various resolutions simultaneously.


In the present paper we adopt a nonparametric model  which is independent of the grid of the forward solver and is reminiscent of non-parametric kernel regression methods. The unknown parametric field is approximated by a superposition of kernel-type functions centered at various locations. The cardinality of the representation, i.e. the number of such kernels,  is treated as an unknown to be inferred in the Bayesian formulation. This gives rise to  a very flexible model that is able to adapt to the problem and the data at hand and find succinct representations of the parametric field of interest. Prior information on the scale of variability can be directly introduced in the model.    

Inference is performed using Sequential Monte Carlo samplers.  
 They utilize  a  set of random samples, named particles, which  are propagated  using simple importance sampling, resampling and updating/rejuvenation mechanisms. 
The algorithm is directly parallelizable as the evolution of each particle is by-and-large independent of the rest. The sequence of distributions defined is based on using solvers that operate on different resolutions and which successively produce  finer discretizations. This results in an efficient hierarchical  approach that makes  use of the results from solvers operating at the coarser scales in order to update them based on analyses on a finer scale. The particulate approximations produced provide concise representations of the posterior which can be readily updated if more data become available or if more accurate solvers are employed.

%% file: definition.tex
In lieu of a formal definition, we discuss  an extremely simple problem which nevertheless  possesses the most important features for the purposes of this work. Consider the  steady-state heat equation in the unit interval, i.e.:
\be
\label{eq:d1}
\frac{d}{dx} \left( -c(x) \frac{dT}{dx} \right) =0, \quad x\in [0,1] 
\ee
where $c(x)$ is the spatially varying conductivity field and $T(x)$ the temperature profile. Assume that known boundary conditions $T(0)=0$  and $\left( -c(x) \frac{dT}{dx} \right)_{x=1}=q$ are imposed and temperature measurements $T_i$ (without any noise) are  obtained  at $N$ distinct points $x_i \in [0,1]$ with the intention of identifying the unknown conductivity and its spatial variation.

For any  interval $\Delta x_i = x_{i+1}-x_i$ between two observation locations, the governing PDE and boundary conditions imply that:
\be
\label{eq:d2}
\left( \int_{x_i}^{x_{i+1}} \frac{1}{c(x)}~dx \right)^{-1}=\frac{q}{T_{i+1}-T_i}
\ee
 
Similar expressions  hold for all other intervals and  relate  the {\em effective} conductivity  in  each subdomain (given by the harmonic mean) with the measured temperature. These relations  however do not {\em uniquely } identify the spatial variability of $c(x)$ unless the latter is assumed constant within each $\Delta x_i$. 
Further constraints can be imposed by assuming continuity of $c(x)$ at the  $x_i$'s but these do not necessarily hold if one considers materials that consist of distinct phases. Even when such constraints seem plausible, one can readily imagine parametric forms of $c(x)$ (i.e. polynomials of high degree) which cannot be completely identified unless  further constraints (e.g. continuity of the derivatives of $c(x)$ at $x_i$) are artificially imposed. The non-uniqueness  persists when the number of measurements $N$ increases even though the space  of possible solutions shrinks.  It also precludes the possibility of detecting significant changes in $c(x)$ that occur in length scales  much smaller than $\Delta x_i$ (e.g. flaws) which are generally  of significance to the analyst. Their contribution to the effective conductivity in \refeq{eq:d2} can be negligible unless  $\Delta x_i$ is of comparable size.
This ill-posedness has long been identified and can become more pronounced in two or three dimensional domains and if the governing PDEs are nonlinear or involve more than one unknown parameters or fields (\cite{kai05com}).
It is also amplified if the measurements obtained are contaminated by random noise which is generally the case in engineering practice. 

Hence there is a need for a general framework that can produce estimates about the unknown fields particularly with regards to the scale of their variability. This is especially important as the  accuracy of the predictions of  computational models is greatly influenced by the  the multiscale nature of property variations. In recent years  a lot of research efforts  have been devoted to the development of scalable, black-box simulators that provide the coarse-scale  solution while capturing the effect of fine-scale  fluctuations (\cite{dol04mul}). The multiscale analysis of such systems inherently assumes that the complete, fine-scale variation of various properties (or model parameters in general) is known. This assumption limits the applicability of these frameworks since it is usually impossible to directly determine   the complete structure of the medium of interest at the finest scale. More often than not, what is experimentally available and accessible (as in the example above), are measurements of the response of these systems under prescribed input or excitation, at spatial scales much coarser than those of the  property variations.                                                        In problems of estimation of soil permeability for example, measurements are restricted to a few bore holes several meters apart from each other. In estimating damage in an aircraft fuselage, measurements of the response (displacements, accelerations etc) are collected at a few locations.  

This limited and noisy information naturally introduces a lot of uncertainty and  necessitates viewing the property variation as a random field whose statistical properties must be consistent with the available data. 
To that end the present paper  proposes a general framework  that is based on the Bayesian paradigm and addresses the following  questions: a) How can one utilize deterministic, forward solvers in order to identify spatial variability of various properties while accounting for the associated uncertainty? b) How can this process produce estimates at various resolutions?, c) As these forward models are computationally demanding, how can this process be done in a computationally efficient manner?, d) How can the available data be used to quantify error or discrepancies in the forward models? 

In the following sections we discuss the characteristics of the proposed Bayesian model with particular emphasis on the prior specifications and their physical implications. We then present a general, efficient inference technique for the determination of the posterior and discuss how predictions in the context of computational models can be achieved. We finally illustrate the capabilities in numerical examples.

%% file: methodology.tex
\subsection{Hierarchical Bayesian Model}

The central goal of this work is to build mathematical methods that utilize limited and noisy observations/measurements in order to identify the spatial variability of model parameters. Given the significant uncertainty arising from the random noise, lack of data and model error, point estimates are of little use. Furthermore it is  important  to quantify the confidence in the estimates made but also in the predictive ability of the  the model of interest.
To that end we adopt a Bayesian perspective. Bayesian formulations  differ from classical statistical approaches (frequentist) in that all unknown parameters (denoted by $\bs{\theta}$) are treated as random. Hence the results of the inference process  are not point estimates but distribution functions. 

The basic elements of Bayesian models are the {\em likelihood} function $L(\bs{\theta})=p(\bs{y} \mid  \bs{\theta})$ which is a conditional probability distribution and gives a (relative) measure of the propensity of observing data $\bs{y}$ for a given model configuration specified by the parameters $\bs{\theta}$. The likelihood function is also encountered in frequentist formulations where the unknown model parameters $\bs{\theta}$ are determined by maximizing $L(\bs{\theta})$. This could be thought as the probabilistic equivalent of deterministic optimization techniques commonly used in inverse problems. It can suffer from the same issues related to the ill-posedeness of the problem.                          The second component of Bayesian formulations  is the {\em prior} distribution $p(\bs{\theta})$ which encapsulates in a probabilistic manner any knowledge/information/insight that is available to the analyst prior to observing the data. Although the prior is a point of frequent criticism due to its inherently subjective nature, it can prove extremely useful in engineering contexts as it provides a mathematically consistent vehicle for  injecting the analyst's insight and physical understanding.                                                             
   The combination of {\em prior} and {\em likelihood} based on Bayes' rule yields the {\em posterior} distribution $\pi(\bs{\theta})$ which probabilistically summarizes the information extracted from the data with regards to the  unknown $\bs{\theta}$ :
\be
\label{eq:m0}
\pi(\bs{\theta})=p( \bs{\theta} \mid \bs{y})=\frac{p(\bs{y} \mid  \bs{\theta})~ p(\bs{\theta}) }{ p(\bs{y}) } \propto p(\bs{y} \mid  \bs{\theta})~ p(\bs{\theta})
\ee                                                                                                                                                             Hence Bayesian formulations allow for the possibility of multiple solutions - in fact any  $\bs{\theta}$ in the support of the likelihood and the  prior is admissible - whose  {\em relative plausibility} is quantified by the posterior. Credible or confidence intervals can be readily estimated from the posterior which quantify inferential uncertainties about the unknowns.


\begin{figure}
\psfrag{data}{data}
\psfrag{y}{$\bs{y}$}
\psfrag{Bayes}{Bayesian Model}
\psfrag{prior}{Nonparametric prior}
\psfrag{s1}{black-box solver}
\psfrag{c}{\underline{coarse resolution}}
\psfrag{s2}{black-box solver}
\psfrag{m}{\underline{medium resolution}}
\psfrag{s3}{black-box solver}
\psfrag{f}{\underline{fine resolution}}
\psfrag{s4}{black-box solver}
\psfrag{finest}{\underline{finest resolution}}
\psfrag{field}{inferred field}
\psfrag{sensors}{\textcolor{red}{*} sensor locations}
\centering
\includegraphics[width=0.95\textwidth]{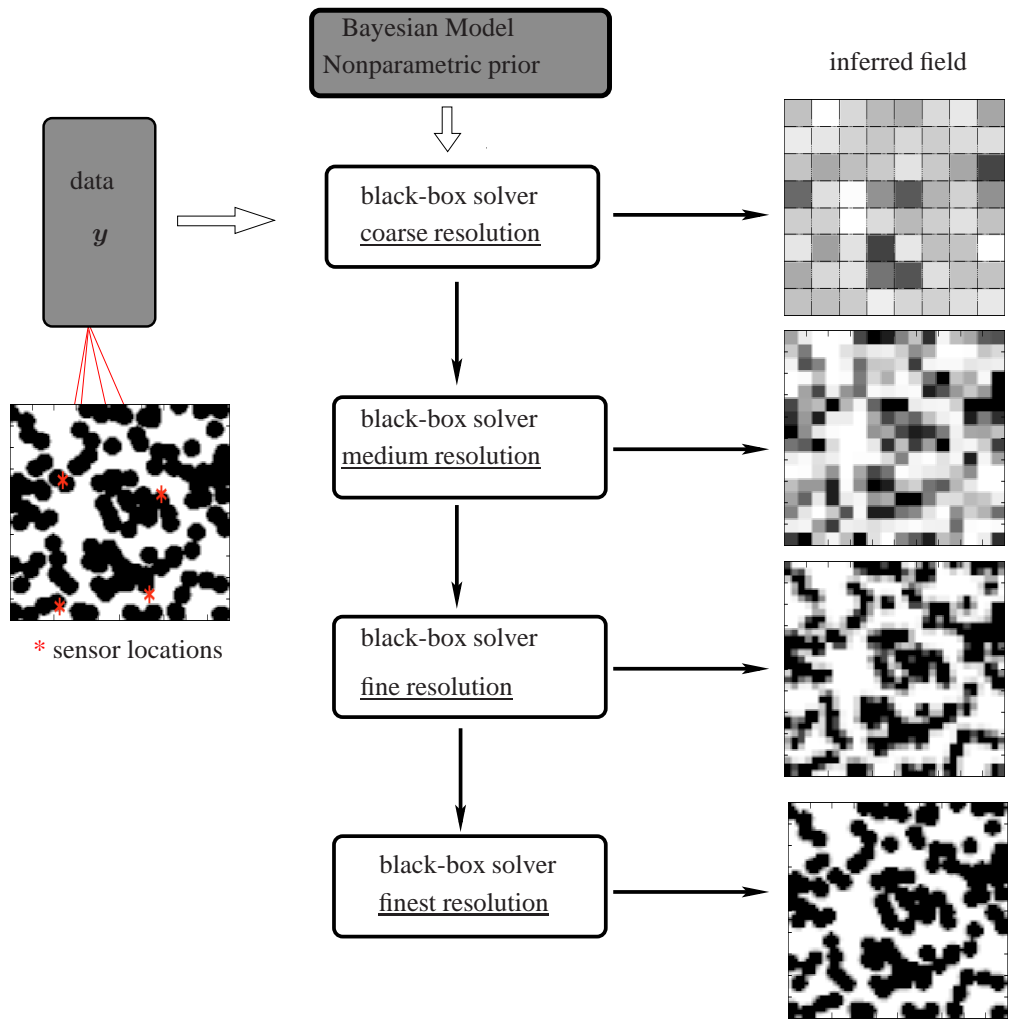}
\caption{Hierarchy of solvers operating on different resolutions  }
\label{fig0}
\end{figure}

Without loss of generality, we postulate the existence of a deterministic, forward model which in most cases of practical interest corresponds to a Finite Element or Finite Difference  model of the governing differential equations. Naturally, forward models allow  for various levels of  discretization of the spatial domain  and let $r$ denote the  resolution they operate upon (larger $r$ implies finer resolution).
In this paper, forward solvers are viewed as {\em messengers}, that carry information about the underlying material properties as they manifest themselves  in the response (mechanical, thermal etc) of the medium of interest.  This is especially true in the context of recently developed  upscaling schemes (\cite{hou97mul,hug98var,dor98wav,kev03equ,e03het,san03cap,aso06sto,kou07sto}) which attempt to capture the effect of finer scale material variability while operating on a coarser grid.
 In general,  the finer the resolution of the forward solver, the more information this provides. This however comes at the expense of computational effort. It is not unusual that the sufficient resolution of the property fluctuations in many systems of practical interest requires  several CPU-hours for a single analysis. Despite the fidelity and accuracy of such high-resolution solvers, they can be of little use in the context of parameter identification as they will generally have to be called upon several times and several system analyses will have to be performed.

Hence an accurate  but expensive {\em messenger} is not the optimal choice if several pieces of information need to be communicated. In many cases however, the fidelity of the message can be compromised if the expense associated with the messenger is smaller. This is especially true if the loss of accuracy can be quantified,  measures of confidence can be provided and furthermore if it leads to the same decisions/predictions.
In this project we propose a consistent framework for using faster but less-accurate forward solvers operating on coarser resolutions in order to expedite property identification. Furthermore these  solvers provide a natural hierarchy of models that if appropriately coupled can further expedite the identification process. Following the analog introduced earlier, we propose using inexpensive messengers (coarse scale solvers), several times to communicate the most pivotal pieces of information and more expensive messengers (fine scale solvers) fewer times to pass on some of the finer details (Figure \ref{fig0}).

 In the remainder of this sub-section, we discuss the basic components of the Bayesian model proposed, with particular emphasis on the prior for the unknown parametric fields. We then present (sub-section \ref{inference}) the proposed inference techniques for the determination of the posterior.

\subsubsection{Likelihood Specification}

Let  $\bs{F}^r=\{F_i^r\}: \mathcal{G} \to \mathcal{E}$ denote the vector-valued  mapping implied by the forward model (operating at resolution $r$), which  given $f(x) \in \mathcal{G}$ (\refeq{eq:0})
 provides the values of response quantities represented by the data $\bs{y}=\{y_i\} \in \mathcal{E}$. This function is the discretized version of the inverse of the differential operator $\mathcal{A}$ in \refeq{eq:0}
 parameterized by $f(\bs{x})$. 
Each evaluation of $\bs{F}^r$ for a specific field  ${f}(\bs{x})$ implies a call  to the  forward  solver (e.g. Finite Elements) that operates on a discretization/resolution $r$. In the proposed framework, the function $\bs{F}^r$ will be treated as a black box.  Naturally data and model predictions will deviate when the former are obtained experimentally due to the unavoidable noise in the measurements.  Most importantly perhaps this deviation can be the result of the model not fully capturing the salient physics either because the governing PDEs are an idealization or because of the discretization error in their solution.
We postulate the following relationship:



\be
\label{eq:m1}
\underbrace{y_i}_{datum ~i} = \underbrace{F_i^{(r)} (f(\bs{x}) )}_{model~prediction} + e_i^{(r)} \quad i=1,2,\ldots,n
\ee
where  $e_i^{(r)}$ quantify the deviation between  model predictions and data, and which will naturally depend on the resolution $r$ of the forward solver.
Quite frequently the data available to us are in the form of disparate observations, that correspond to different physical phenomena (e.g. temperatures and displacements in a thermo-mechanical problem) in which case the computational model corresponds to a coupled multiphysics  solver.

The probabilistic model for $e_i^r$ in  \refeq{eq:m1} gives rise to the {\em likelihood function} (\refeq{eq:m0}). In the simplest case where $e_i^{(r)}$  are assumed independent, normal variates with zero mean and variance $\sigma_r^2$:
\vspace{-.5cm}
\begin{eqnarray}
\label{eq:m9b}
p_r(y_i \mid {f}(x), \sigma_r) & \propto & \frac{1}{\sigma_r}\exp \{-\frac{1}{2} \frac{ \left(y_i-F_i^{(r)} (f(\bs{x}) ) \right)^2}{\sigma_r^2} \}  \nonumber \\
 \textrm{and  } ~ p_r(\bs{y} \mid {f}(x), \sigma_r)& \propto &  \frac{1}{\sigma_r^n} \exp \{-\frac{1}{2\sigma_r^2} \sum_{i=1}^n \left(y_i-F_i^{(r)} (f(\bs{x}) ) \right)^2 \}
\vspace{-.3cm}
\end{eqnarray}
More complex models which can account for the spatial dependence of the error variance $\sigma_r^2$ or the detection of events associated with sensor malfunctions at  certain locations, can readily be formulated. 
In general the variances  $\sigma_r^2$ are unknown (particularly the component that pertains to model error) and should be inferred from the data. When a conjugate,  $Gamma(a,b)$ prior is adopted for $\sigma_{r}^{-2}$, the error variances can be integrated out from \refeq{eq:m9b} further simplifying the likelihood:
\vspace{-.3cm}
\be
\label{eq:m9c}
L_r({f}(x))=  p(\bs{y} \mid {f}(x))  \propto   \frac{\Gamma (a+n/2) }{ \left( b+\frac{1}{2}  \sum_{i=1}^n \left(y_i-F_i^{(r)} (f(\bs{x}) \right)^2 \right)^{a+n/2} }
\ee
where $\Gamma(z)=\int_0^{+\infty} t^{z-1} ~e^{-t}~dt$ is the gamma function.

It should be noted that in some works (\cite{Kennedy:2001,Higdon:2008}), explicit distinction between model and observation errors is made, postulating a relation of the following form:
\be
observation/data = model~ prediction~+~model~ error~+~observation~error
\ee
 As it has been observed (\cite{Wynn:2001}), independently of the amount of data available to us, these three components are not {\em identifiable}, meaning several different values can be equally consistent with the data. This however does not imply that all possible values are equally plausible. For example a large number of  values of the observation error  that are all positive or all negative (for all observations) are not consistent with the perception of random noise but most likely imply a bias of the model or perhaps a miscalibrated sensors used to collect the data.  Bayesian formulations are highly suited for such problems as they provide a natural way of quantifying a priori and a posteriori relative measures of plausibility. In the following we restrict the presentation  on models of \refeq{eq:m1} as  the focus of  is on identifying the scale of variability of material properties ${f}(\bs{x})$. 


\subsubsection{Prior  Specification}
\label{prior}
The most critical component involves the prior specification for the unknown material properties as represented by ${f}(x)$.
 In existing Bayesian (\cite{Wang:2005a,joh05mul}), but also deterministic (optimization-based), formulations, ${f}(x)$ is discretized according to the spatial resolution of the forward solver.  For example, in  cases where  finite elements are used, the property of interest is assumed constant within each element and therefore the vector of unknowns is of  dimension equal to the number of elements.
This offers obvious implementation advantages  but also poses some difficulties since the scale of variability  of material properties is implicitly selected by the solver rather than the data.
 This is problematic in several ways. On one hand if the scale of variability is larger than the grid, a waste of resources takes place, at the solver level which has to be run at unnecessarily fine resolutions, and at the level  of the inference process which is  impeded by the unnecessarily large dimension of the  vector of unknowns.  Furthermore, as the number of unknowns is much larger by comparison to the amount of  data it can lead to {\em over-fitting}.  This will produce erroneous or even absurd values for the unknowns that may nevertheless fit perfectly the data. Such solutions  will have negligible  {\em predictive ability} and would be useless in decision making.
 On the other hand, if the scale of variability is smaller than the grid, it cannot be identified even if the solver provides sufficient information for discovering this possibility.

 In order to increase the flexibility of the model, we base our prior models for the unknown field(s) $f(\bs{x})$   on the convolution representation of a Gaussian process. 
      An alternative representation of a stationary Gaussian process involves a convolution of a  white noise process $a(\bs{x})$   with a smoothing kernel $K(. ; \phi)$ depending on a set of parameters $\phi$ (\cite{bar96bla,hig02spa}):
\be
\label{eq:conv1}
f(\bs{x}) = \int K(\bs{x}-\bs{z}; \phi) ~a(\bs{z})~d\bs{z}
\ee
The kernel form determines essentially the covariance of the resulting process, since:
\be
\label{eq:conv2}
cov \left( f(\bs{x}_1, f(\bs{x}_2)\right)= E[f(\bs{x}_1, f(\bs{x}_2)]=\int K(\bs{x}_1-\bs{z}; \phi)  K(\bs{x}_1-\bs{z}; \phi) ~ d\bs{z}
\ee
For computational purposes, a discretized version of \refeq{eq:conv1} is used:
\be
\label{eq:conv3}
{f}(\bs{x}) = \sum_{j=1}^k a(\bs{z}_j) K(\bs{x}-\bs{z}_j; \phi) = \sum_{j=1}^k a_j K(\bs{x}-\bs{x}_j; \phi)
\ee

In order to increase the expressive ability of the aforementioned model we introduce two improvements. Firstly we consider that the set of kernel parameters $\phi$ is spatially varying resulting in a non-stationary process:
\vspace{-.3cm}
\be
\label{eq:m2}
{f}(\bs{x}) = a_0 + \sum_{j=1}^k a_j K_j(\bs{x}; \bs{\phi}_j) \quad x \in D
\ee 
where $a_0$ corresponds to  a value of $\phi_0$ such that  the corresponding kernel is $1$ everywhere.
Such representations can be viewed  as a radial basis network as in \cite{tip01spa}). Furthermore by interpreting the kernels as basis functions, \refeq{eq:m2} it can be seen as an extension of the  the representer theorem of Kimeldorf and Wahba (\cite{kim71cor}).  Overcomplete representations as in \refeq{eq:m2} have
been advocated because they have greater robustness in the presence of 
noise, can be sparser, and can have greater flexibility in matching structure
in the data (\cite{lew00lea,lia06non}).  One possible selection for the functional form of $K_j$, that also has an  intuitive parameterization with regards to the scale of of variability in the material properties, is isotropic, Gaussian kernels:
\be
\label{eq:m3}
K( \bs{x} ; \bs{\phi}_j=(\bs{x}_j, \tau_j)) = \exp \{ - \tau_{j} \parallel \bs{x} -\bs{x}_j \parallel^2 \}
\ee
The parameters $\tau_j$  directly correspond to the scale of variability of ${f}(\bs{x})$. Large $\tau_j$'s imply narrowly concentrated  fluctuations and large values slower varying fields. The center of each kernel is specified by the location parameter $\bs{x}_j$. Other functional forms (e.g. discontinuous) can also be used on their own or in combinations to enrich the expressivity of the expansion in \refeq{eq:m2}. 
Wavelets, steerable wavelets, segmented wavelets, Gabor dictionaries, multiscale Gabor dictionaries, wavelet packets, cosine packets, chirplets, warplets, and a wide range of other dictionaries that have been developed in various contexts (\cite{che01ato}) offer several possibilities.

The second important improvement is that we allow the size of the expansion $k$ to vary. It is obvious that such an assumption is consistent with the {\em principle of parsimony},  which states that prior models should make as few assumptions as possible and allow their complexity to be inferred from the data.  
Hence  the {\em cardinality } of the model, i.e. the number of basis functions $k$ is the key unknown that must be determined so as to  provide a good interpretation of the observables.

Independently of the form of the kernel adopted, the important, common characteristic of all such approximations (as in \refeq{eq:m2}) is that the field representation {\em does not depend on the resolution of the forward model}. The latter  affects inference only through the black-box functions $F_i^r$ (\refeq{eq:m1}, Figure \ref{fig0})) as it will be illustrated in the next sections.

The parameters of the prior model adopted consist of:
\bi
\item $k$: the number of kernel functions needed, 
\item $\{a_j\}_{j=1}^k$, the coefficients of the expansion  in \refeq{eq:m2}. Each of those can be a scalar or vector depending on the number of material property fields we want to infer simultaneously. For example, in a  problem of thermo-mechanical coupling where the data consists of temperatures and displacements and we want to identify elastic modulus and conductivity, each $a_j$ will be a vector in $\RR^2$.
\item $\{ \tau_j \}_{j=1}^k$ the precision parameters of each kernel which pertain to the scale of the unknown field(s), and 
\item $\{ \bs{x}_j \}_{j=1}^k$ the locations of the kernels which are points in $\mathcal{D}$.
\ei

In accordance with the Bayesian paradigm, all unknowns are considered random and are assigned  prior distributions which quantify any information, knowledge, physical insight, mathematical constraints  that is available to the analyst before the data is processed. Naturally, if specific prior information is available it can be reflected on the prior distributions. We consider prior distributions of the following form (excluding hyperparameters):
\begin{eqnarray}
\label{eq:m10}
p(k, \{a_j\}_{j=0}^k,\{\tau_j\}_{j=1}^k, \{\bs{x}_j\}_{j=1}^k) & \propto &  p(k)   \nonumber \\ 
& \times & p( \{a_j\}_{j=0}^k \mid k )  \nonumber \\
& \times &  p(\{\tau_j\}_{j=1}^k \mid k)  \nonumber \\
& \times & p( \{\bs{x}_j\}_{j=1}^k) )    
\end{eqnarray}

 In order to increase the robustness of the model and exploit structural dependence we adopt a hierarchical prior model (\cite{gel03bay}).

\vspace{.2cm}
\noindent {\bf Model Size:} 

Pivotal to the robustness and expressivity  of the model  is the selection of the model size, i.e. of the number of kernel functions $k$ in \refeq{eq:m2}. This number is unknown a priori and in the absence of specific information, {\em sparse} representations should be favored.  This is not only advantageous for computational purposes, as the number of unknown parameters is proportional to $k$, but also consistent with the parsimony of explanation principle or Occam's razor (\cite{jef92ock,ras01occ,mur05not}). For that purpose, we propose  a truncated Poisson prior for $k$:
\be
\label{eq:m5}
p(k \mid \lambda) \propto \left\{ \begin{array}{cc} e^{-\lambda} \frac{\lambda^k}{k!} & \textrm{if } k \le  k_{max} \\ 0 & \textrm{otherwise} \\ \end{array} \right.
\ee
The truncation parameter $k_{max}$ is selected based on computer memory limitations and defines the support of the prior. This prior allows for representations of various cardinalities to be assessed  simultaneously with respect to the data. As a result the number of unknowns is not  fixed and the corresponding posterior has support on spaces of different dimensions as discussed in more detail in the sequence. 
 In this work, an exponential  hyper-prior is used for the hyper-parameter $\lambda$ to allow for greater flexibility and robustness i.e. $p(\lambda \mid s) =s~\exp\{-\lambda ~s \}$. After integrating out $\lambda$ we obtain:
\be
\label{eq:m5a}
p(k \mid s) \propto \frac{1}{(s+1)^{k+1}} ,  \qquad for ~ k=0,1,\ldots, k_{max}
\ee

\noindent {\bf Scale:}

The most critical perhaps parameters of the model are $\{ \tau_j\}_{j=1}^k$ which control the {\em scale of variability} in the approximation of the unknown field(s). If prior information about this is available then it can be readily accounted for by appropriate prior specification. In the absence of such information however  multiple possibilities exist. In contrast to deterministic optimization techniques where ad-hoc {\em regularization} assumptions are made, in the Bayesian framework proposed possible solutions are evaluated with respect to their {\em plausibility} as quantified by the posterior distribution. This provides a unified interpretation of various assumptions that are made regarding the priors of the parameters involved.
 For example, consider a  general $Gamma(a_{\tau},b_{\tau})$ prior:
\be
\label{eq:m9}
p(\{\tau_j\}_{j=1}^k \mid k, ~a_{\tau},b_{\tau} ) = \prod_{j=1}^k \frac{b_{\tau}^{a_{\tau}} }{\Gamma(a_{\tau}) } \tau_j^{ a_{\tau}-1 } ~\exp(-b_{\tau} \tau_j)
\ee
This has a mean $a_{\tau} / b_{\tau}$ and coefficient of variation $1/\sqrt{a_{\tau}}$. Diffuse versions  can be adopted  by selecting small $a_{\tau}$.
 A non-informative prior $p(\tau_j)\propto 1/\tau_j $ arises as a special case  for $a_{\tau}=2$ and $b_{\tau}=0$ which is invariant under rescaling. Furthermore. it offers an interesting physical interpretation as it favors ``slower'' varying representations (i.e. smaller $\tau$'s). 
In order to automatically determine the mean of the Gamma prior, we express  $b_{\tau}=\mu_{j} a_{\tau} $ where $\mu_j$ is a location parameter for which  an Exponential hyper-prior is used with a hyper-parameter $a_{\mu}$ i.e. $p(\mu_j) =\frac{1}{a_{\mu}}  e^{-\mu_j/a_{\mu}}$. Integrating out the $\mu_j$'s leads to following prior:
\be
\label{eq:m99}
p(\{\tau_j\}_{j=1}^k \mid k, ~a_{\tau},a_{\mu} ) = \prod_{j=1}^k     \frac{\Gamma(a_{\tau}+1) }{ \Gamma(a_{\tau}) }~ a_{\tau}^{a_{\tau}} ~\frac{\tau_j^{(a_{\tau}-1)}   }{a_{\mu}}~\frac{1}{(a_{\tau}\tau_j+a_{\mu}^{-1})^{(a_{\tau}+1)} }  
\ee


\noindent {\bf Other Parameters:}

For the coefficients $a_j$ a multivariate normal prior was adopted:
\be
\label{eq:m6}
 \{{a}_j\}_{j=0}^k \mid k, \sigma_a^2 \sim N(\bs{0}, \sigma_a^2~\bs{I}_{k+1} )
\ee
where $\bs{I}_{k+1}$ is the $(k+1)\times(k+1)$ identity matrix. 
The hyper-parameter $\sigma^2_a$ which controls the spread of the prior is modeled by the standard inverse gamma distribution $ Inv-Gamma(a_0,b_0)$. It can readily be integrated-out leading to the following prior for $a_j$'s:
\be
\label{eq:m6a}
p(\{{a}_j\}_{j=0}^k \mid k, ~a_0,b_0)=\frac{1}{(2\pi)^{(k+1)/2} } \frac{\Gamma(a_0+\frac{k+1}{2} ) }{ \left( b_0+\frac{1}{2} \sum_{j=0}^k a_j^2 \right)^{a_0+(k+1)/2} }
\ee


Finally, for the unknown kernel locations $\{\bs{x}_j\}_{j=1}^k$, a uniform prior in $\mathcal{D}$ is proposed i.e.:
\be
\label{eq:m8}
p(\{\bs{x}_j\}_{j=1}^k \mid k) = \frac{1}{\mid \mathcal{D} \mid^k}
\ee 
where $\mid \mathcal{D} \mid$ is the length or area or volume  of $\mathcal{D}$ in one, two or three dimensions respectively. Naturally if prior information is available about  subregions with significant property variations this can be incorporated in the prior.

\noindent {\bf Complete Model:} 

Let $\bs{\theta}_k=\{\{a_j\}_{j=0}^k,\{ \tau_j \}_{j=1}^k, \{ \bs{x}_j \}_{j=1}^k\}\in \bs{\Theta}_k$  denote the vector containing all the unknown  parameters and $\bs{\theta}=(k,\bs{\theta}_k)$. Since $k$ is also assumed unknown and allowed to vary, the dimension of $\bs{\theta}_k$ is variable as well and $\bs{\Theta}_k\triangleq (\RR^{k+1} \times (\RR^+)^k \times \mathcal{D}^k$.
 In 2D for example and assuming a scalar unknown field $f(x)$ in the expansion of \refeq{eq:m2} the dimension of $\bs{\theta}_k$ is $(k+1)+k+2k=2+4k$. 
Based on \refeq{eq:m10} and Equations (\ref{eq:m5a}), (\ref{eq:m9}), (\ref{eq:m6a}) and (\ref{eq:m8}), the complete prior model is given by:
\begin{eqnarray}
\label{eq:prior}
p(\bs{\theta} \mid s,~a_{\tau},a_{\mu},~a_0,b_0) & = & \frac{1}{(s+1)^{k+1}} \nonumber \\
 & \times  & \prod_{j=1}^k     \frac{\Gamma(a_{\tau}+1) }{ \Gamma(a_{\tau}) }~\frac{ a_{\tau}^{a_{\tau}} }{\tau_j^{(a_{\tau}-1)} }~\frac{1}{a_{\mu}}~\frac{1}{(a_{\tau}\tau_j+a_{\mu}^{-1})^{(a_{\tau}+1)} }  \nonumber \\
 & \times & \frac{1}{(2\pi)^{(k+1)/2} } \frac{\Gamma(a_0+\frac{k+1}{2} ) }{ \left( b_0+\frac{1}{2} \sum_{j=0}^k a_j^2 \right)^{a_0+(k+1)/2} } \nonumber \\
& \times & \frac{1}{\mid \mathcal{D} \mid^k}
\end{eqnarray}

The combination of the prior   $p(\bs{\theta})$ with the likelihood $L_r(\bs{\theta})$ (\refeq{eq:m9c}) corresponding to a forward solver operating on resolution $r$, give rise to the {\em posterior} density $\pi_r(\bs{\theta})$ which is proportional to:
\be
\label{eq:m13}
\pi_r(\bs{\theta})=p_r(\bs{\theta} \mid \bs{y}) \propto L_r(\bs{\theta}) ~p(\bs{\theta}) 
\ee

Even though several parameters have been removed from the vector of unknowns $\bs{\theta}$ and marginalized in the pertinent expressions, the corresponding posteriors can be readily be obtained, or rather be sampled from, once the posteriors $\pi_r(\bs{\theta})$ have been determined. As it is shown in the numerical examples, of interest could be the variance $\sigma_r^2$ of the error term (Equations (\ref{eq:m1}), (\ref{eq:m9b})) which quantifies the magnitude of the deviation between model and data and can serve as a validation metric (in the absence of observation error)  or be used for predictive purposes  (see section \ref{prediction}). From \refeq{eq:m1} and the conjugate prior model adopted for $\sigma_r^2$, it can readily be shown that the conditional posterior is given by a Gamma distribution:
\begin{eqnarray}                                                                              \label{eq:m12}
p(\sigma_r^{-2} ,\bs{\theta} \mid  \bs{y}) & = & p(\sigma_r^{-2} \mid \bs{\theta} ) ~ \pi_r(\bs{\theta} \mid  \bs{y}) \nonumber \\
 & \textrm{and} & \nonumber \\
p(\sigma_r^{-2} \mid \bs{\theta} ) & = & Gamma \left( a+\frac{n}{2}, b+ \frac{ \sum_{i=1}^n \left(y_i-F_i^{(r)} (\bs{\theta}) \right)^2 }{ 2 } \right)
\end{eqnarray}
In the context of Monte Carlo simulation, this trivially implies that once samples $\bs{\theta}$ from $\pi_r$ have been obtained, the samples of $\sigma_r^{-2}$ can also be drawn from the aforementioned Gamma.

The support of the posteriors $\pi_r$  lies  on $\cup_{k=0}^{k_{max}}\{k\} \times  \bs{\Theta}_k$. Two important points are worth emphasizing. Firstly, \refeq{eq:m13} defines a {\em sequence of posterior densities}, each corresponding to a different likelihood and a different forward solver of resolution $r$. It is clear that the black-box functions  $\bs{F}^{(r)}$ appearing in the likelihood in  \refeq{eq:m9b} imply {\em denser} mappings for smaller $r$. This is because  solvers corresponding to coarser resolutions of the governing PDEs are more myopic (compared to solvers at finer resolutions) to small scale fluctuations of the spatially varying model parameters ${f}(\bs{x})$ (parameterized by $\bs{\theta}$). As a result the likelihood functions $L_r$ and the associated posteriors $\pi_r$ will be  flatter and have fewer modes for smaller $r$. The task of identifying these posteriors becomes increasingly more difficult as we move to solvers of higher refinement (i.e. larger $r$). It is this feature that we propose of exploiting in the next section in order to increase the accuracy and improve on the efficiency of the inference process. 
In addition,  the posteriors $\pi_r$ are only known up to a normalizing constant (determining $p(\bs{y})$ in \refeq{eq:m0} involves an infeasible and unnecessary  integration in a very high dimensional space). Each evaluation of $\pi_r$ for a particular $\bs{\theta}$ requires calculating  $\bs{F}^{(r)}$ and therefore calling the corresponding black-box solver. As each of these runs of the forward solver may involve the solution of very large systems of equations they can be extremely time consuming.  It is important therefore to determine $\pi_r$ not only accurately, but also with the least possible number of calls to the forward solver. Since solvers corresponding to coarser resolutions (smaller $r$) are faster, it would be desirable to utilize the information they provide in order to reduce the number of calls to more expensive, finer resolution solvers.  

\subsection{Determining the Posterior - Inference }
\label{inference}

The posterior defined above is analytically intractable. For that reason, {\em Monte Carlo} methods provide essentially the only accurate way to infer $\pi_r$. Traditionally {\em Markov Chain Monte Carlo} techniques (MCMC) have been employed to carry out this task (\cite{Higdon:2002,Lee:2002,Lee:2002a,Wang:2004,lee07mul}). These are based on  building a Markov chain that asymptotically converges to the target density (in this case $\pi_r$) by appropriately defining a transition kernel.  While convergence can be assured under weak conditions (\cite{liu01mon,rob04mon}), the rate of convergence can be extremely slow and require a lot of likelihood evaluations and calls to the black-box solver. Particularly in cases where the target posterior can have multiple modes, very large {\em mixing times} might be required which constitute the method impractical or infeasible. In addition, MCMC is not directly parallelizable, unless multiple independent chains are run simultaneously and it can be difficult to design a good proposal distribution when operating in high dimensional spaces.
 More importantly perhaps, standard MCMC is not capable of providing a {\em hierarchical, multi-resolution} solution to the problem. Consider for example, the case that several samples have been drawn using MCMC from  the posterior $\pi_{r_1}$ corresponding to  a solver operating on  resolution $r=r_1$. If samples of  the posterior $\pi_{r_2}$ are needed, corresponding to a solver of finer resolution $r_2 > r_1$ but not significantly different from $r_1$, then MCMC iterations would have to be {\em initiated  anew}. Hence there is no immediate way to exploit the inferences made about $\pi_{r_1}$ even though the latter might be quite similar to $\pi_{r_2}$.

In this work we advocate the use of {\em Sequential Monte Carlo} techniques (SMC).  They
represent  a set of flexible simulation-based methods for sampling from a sequence of probability distributions (\cite{mac98seq,dou01seq}).  As with Markov Chain Monte Carlo methods (MCMC), the target
distribution(s) need only be known up to a constant and therefore
 do not require calculation of the intractable integral in the denominator in \refeq{eq:m0}.  
 They utilize  a  set of random samples (commonly referred to as {\em particles}), which  are
propagated  using a combination of {\em importance sampling},  {\em resampling}  and MCMC-based {\em rejuvenation} mechanisms (\cite{mor06seq,del06seq}). Each of these particles, which  can be thought of as a possible configuration of the system's state, is associated with an {\em importance  weight} which is proportional to the the posterior value of the respective particle. 
These weights are updated sequentially along with the particle locations. Hence if $\{\bs{\theta}_r^{(i)},~w_r^{(i)} \}_{i=1}^N$ represent $N$ such particles and associated weights for distribution $\pi_r(\bs{\theta})$ then:
\be
\label{eq:m14}
\pi_r(\bs{\theta}) \approx \sum_{i=1}^N ~W_r^{(i)}~ \delta_{\bs{\theta}_r^{(i)}} (\bs{\theta})
\ee
where $W_r^{(i)}=w_r^{(i)}/\sum_{i=1}^N w_r^{(i)}$ are the normalized weights and  $\delta_{\bs{\theta}^{(i)}_r }(.)$ is the Dirac function centered at $\bs{\theta}_r^{(i)}$. Furthermore, for any function $h(\bs{\theta})$  which is $\pi_r$-integrable (\cite{mor04fey,Chopin:2004}):
\be
\label{eq:m15}
 \sum_{i=1}^N ~W_r^{(i)}~h(\bs{\theta}_r^{(i)})  \rightarrow \int h(\bs{\theta})~\pi_r(\bs{\theta})~d\bs{\theta} \quad \textrm{almost surely}
\ee
\noindent Before discussing the SMC sampler proposed, it is worth recapitulating the basic desiderata:
\bi
\item[a)] Accuracy: the Monte Carlo scheme should be able to correctly sample from {\em multi-modal} distributions
\item[b)] Hierarchical, Multiscale: the Monte Carlo scheme should be able to exploit inferences made using forward solvers corresponding to coarser resolutions and refine them as more elaborate forward solvers are used.
\item[c)] Efficiency: the Monte Carlo sampler should require the fewest possible calls to the forward solver. It should be  directly parallelizable and utilize inferences made using cheaper forward solvers corresponding to coarser resolutions in order to reduce the number of calls to more expensive forward solvers corresponding to finer resolutions.
\ei

\begin{figure}
\psfrag{p0}{$\pi_1(\bs{\theta})$}
\psfrag{p1}{$\pi_2(\bs{\theta})$}
\psfrag{p2}{$\pi_{12,\gamma_{s_1}}(\bs{\theta})$}
\psfrag{p3}{$\pi_{12,\gamma_{s_2}}(\bs{\theta})$}
\psfrag{coarse}{coarse}
\psfrag{fine}{fine}
\psfrag{bridge}{bridging scales}
\psfrag{t}{$\bs{\theta}$}
\psfrag{g0}{$\gamma_0=0$}
\psfrag{g1}{$\gamma_S=1$}
\psfrag{g2}{$\gamma_{s_1}=0.25$}
\psfrag{g3}{$\gamma_{s_2}=0.75$}
\includegraphics[width=0.95\textwidth,height=7.5cm]{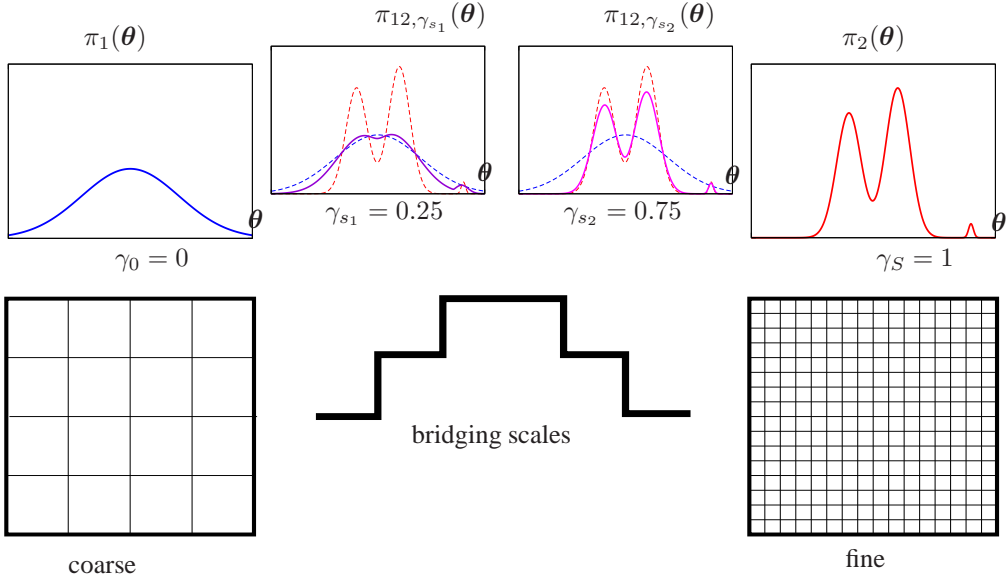}
\caption{Illustration of bridging densities as defined in \refeq{eq:m16} between posterior distributions $\pi_1(\bs{\theta})$, $\pi_2(\bs{\theta})$ corresponding to different resolutions of the governing PDEs. These allow for accurate and computationally efficient transmission of the inferences made to finer scales. }
\label{fig1}
\vspace{.5cm}
\end{figure}

The goal is to obtain samples from each of the posterior distributions in \refeq{eq:m13} corresponding to solvers with increasingly finer spatial resolution of the governing PDEs, $r=r_1,r_2, \ldots, r_M$ where $r_1$  is the coarsest to $r_M$ the finest. For economy of notation we define the artificial posterior $\pi_{r_0}(\bs{\theta})=p(\bs{\theta})$  that coincides with the prior (which is common to all resolutions and independent of the forward solver). To demonstrate the proposed process it suffices to consider a pair of these posterior densities $\pi_{1}(\bs{\theta}) \propto L_1(\bs{\theta})~p(\bs{\theta})$ and $\pi_{2}(\bs{\theta}) \propto L_2(\bs{\theta})~p(\bs{\theta})$ corresponding to forward solvers at two successive resolutions $r_{i_1}$ and $r_{i_2}$ (Figure \ref{fig1}) and discuss the inferential transitions.  Let $\pi_{12,\gamma}(\bs{\theta})$ denote a sequence of artificial, auxiliary  distributions defined as follows:
\be
\label{eq:m16}
\pi_{12,\gamma}(\bs{\theta})=\pi_1^{(1-\gamma)}(\bs{\theta}) ~\pi_2^{\gamma}(\bs{\theta})=L_1^{(1-\gamma)}(\bs{\theta})~L_2^{\gamma} (\bs{\theta})~p(\bs{\theta}) \qquad \gamma \in [0,1]
\ee
where $\gamma$ plays the role of {\em reciprocal temperature}. Trivially for $\gamma=0$ we recover $\pi_1$ and for $\gamma=1$, $\pi_2$. The role of these auxiliary distributions is to {\em bridge the gap  between $\pi_1$ and $\pi_2$} and provide a smooth transition path where importance sampling can be efficiently applied. In this process, inferences from the coarser scale solver are  {\em transferred  and updated} to conform with the finer scale solver. Starting with a particulate approximation for $\pi_{r_0}(\bs{\theta})=p(\bs{\theta})$ (which trivially involves  drawing samples from the prior with weights $w_0^{(i)}=1$), the goal is to gradually update the importance weights and particle locations in order to approximate the target posteriors at various resolutions. In order to implement computationally  such  a transition we define an increasing sequence of $\{\gamma_s\}_{s=1}^S$  with $\gamma_0=0$ and $\gamma_S=1$ (see sub-section \ref{bridging}).  An SMC-based inference scheme would then proceed as described in  Table \ref{tab:smc}.

\begin{table}[htp!]
\begin{tabular}{|p{12cm}|}
\hline
\noindent {\bf SMC algorithm:}
\begin{enumerate}
\item For $s=0$, let $\{\bs{\theta}_0^{(i)},~w_0^{(i)} \}_{i=1}^N$ be the initial  particulate approximation to $\pi_{12,\gamma_{0}}=\pi_1$. Set $s=1$.  

\item {\em Reweigh}: Update weights $w_{s}^{(i)} =w_{s-1}^{(i)}~\frac{ \pi_{12,\gamma_{s}}(\bs{\theta}_{s-1}^{(i)}) }{ \pi_{12,\gamma_{s-1}}( \bs{\theta}_{s-1}^{(i)}) }$

\item {\em Rejuvenate}: Use an MCMC kernel $P_s(.,.)$ that leaves $\pi_{12,\gamma_s}$ invariant to perturb each particle $\bs{\theta}_{s-1}^{(i)} \to \bs{\theta}_{s}^{(i)}$  

\item {\em Resample}: Evaluate the Effective Sample Size,  $ESS=1/\sum_{i=1}^N (W_{s+1}^{(i)})^2$  and resample the population if it is less than a prescribed threshold $ESS_{min}$.

\item The current population $\{\bs{\theta}_{s}^{(i)}, w_{s}^{(i)} \}_{i=1}^N$ provides a particulate approximation of $\pi_{12,\gamma_s}$  in the sense of Equations (\ref{eq:m14}), (\ref{eq:m15}).

\item If $s < S$ (and $\gamma_s < 1$) then set $s=s+1$ and goto to step 2. Otherwise stop.
\end{enumerate} \\
\hline
\end{tabular}
\caption{Basic steps of an  SMC algorithm }
\label{tab:smc}
\end{table}

\noindent{\bf Notes:}
\bi 

\item The role of the {\em Reweighing} step is to correct for  the discrepancy between the two successive distributions in exactly the same manner that importance sampling is employed. The {\em Resampling} step aims at reducing the variance of the particulate approximation by eliminating particles with small weights and multiplying the ones with larger weights.
The metric that we use in carrying out this task is the Effective Sample Size (ESS, Table \ref{tab:smc}) which provides a measure of degeneracy in the population of particles as quantified by their variance. If this degeneracy exceeds a specified threshold,  resampling is performed. As it has been pointed out in several studies (\cite{dou06eff}), frequent resampling can  deplete the population of its informational content and result in particulate approximations that consist of even a single particle. Throughout  this work $ESS_{min}=N/2$ was used.
 Although other options are available, {\em multinomial} resampling is most often applied and was found sufficient in the problems examined.

 \item A critical component  involves the perturbation of the population of samples by a standard MCMC kernel in the {\em Rejuvenation} step as this determines how fast the transition takes place. Although there is  freedom in selecting the transition kernel $P_s(.,.)$ (the only requirement is that it is $\pi_{12,\gamma_s}$-invariant), there is a  distinguishing feature that will be elaborated further in the next sub-section (see \ref{rjmcmc}).  The target posteriors $\pi_r$ (as well as the intermediate bridging distributions in \refeq{eq:m16}) live in spaces of varying dimensions as previously discussed. Hence an exploration of the state space must involve {\em trans-dimensional } proposals. Pairs of such moves can be defined in the context of Reversible-Jump MCMC (RJMCMC , \cite{Green:1995}) such as {\em adding/deleting} a kernel in the expansion of \refeq{eq:m2}, or {\em splitting/merging} kernels (see \ref{rjmcmc}). Even though it is straightforward to satisfy the invariance constraint in the RJMCMC framework, it is more difficult to design moves that also mix fast. As each (RJ)MCMC requires a likelihood evaluation and a call to a potentially  expensive forward solver, it is desirable to minimize their number while retaining good convergence properties. 

\item In most implementations of such SMC schemes, the sequence of intermediate, bridging distributions  is fixed a priori. In order to ensure a smooth transition, a large number is selected at very closely spaced  $\gamma_s$.  It is easily understood that for reasons of computational efficiency, it is desirable to minimize the number of intermediate bridging distributions while ensuring that the successive distributions are not significantly different. In sub-section (\ref{bridging}) we discuss a novel adaptive scheme that allow the automatic determination of these distributions resulting in significant computational savings. 

\item It should  be noted that the framework proposed is directly {\em parallelizable}, as the evolution (reweighing, rejuvenation) of each particle is {\em independent} of the rest. Hence the computational effort can be readily distributed to several processors. 

\item The particulate approximations obtained at each step, provide a {\em concise } summary of the posterior distribution based on the respective forward solver. This can be readily updated in the manner explained above, if forward solvers at finer resolutions become available or computationally feasible. {\em Similar bridging distributions can be established between distinct forward solvers with differences going beyond their respective resolutions}. This is made possible by the {\em  nonparametric Bayesian model}  which is independent of the forward solver and the {\em flexible inference engine based on SMC}.

\item An advantageous feature of the proposed framework is that the confidence in the estimates made can be readily quantified by establishing posterior (or credible) intervals, i.e. the posterior probability that the unknown field of interest $f(x)$ exceeds or not a specified threshold, from the particulate approximations (\refeq{eq:m14}). It is these credible intervals (or in general measures of the variability in the estimates such as the posterior variance) that can guide {\em adaptive refinement} of the governing PDEs. Traditionally, adaptive refinement has been based  on  estimates of some  error norm in the solution of the governing PDEs (\cite{ain00pos}).  This however is inefficient and inadequate for the purposes of identifying spatially varying model parameters as solution errors are not necessarily correlated with the confidence in the estimates. 
It is envisioned that the posterior variance at each point $\bs{x} \in \mathcal{D}$ in the domain interest can serve as the basis for increasing the resolution of the solver at select regions and making optimal use of the computational resources available.
\ei

\subsubsection{Bridging distributions $\pi_{12,\gamma_s}$}
\label{bridging}

The role of these auxiliary distributions is to facilitate the transition between two different posteriors $\pi_1$ and $\pi_2$ corresponding to two distinct solvers. 
It is easily understood that if  $\pi_1$ and $\pi_2$ are not significantly different, then fewer bridging distributions will be needed and vice versa. As it is impossible   to know  a priori  how pronounced these  differences are, in most implementations  a rather large number of bridging distributions is adopted, erring on the side of safety.
We propose an adaptive SMC algorithm, that extends existing versions (\cite{del06seq,mor06seq}) in that it automatically determines the number of intermediate bridging distributions needed. In this process we are guided by the   Effective Sample Size (ESS, Table \ref{tab:smc}) which  provides a measure of degeneracy in the population of particles. If $ESS_s$ is the $ESS$ of the population after the step $s$  and in the most favorable scenario that the next bridging distribution $\pi_{12,\gamma_{s+1}}$ is very similar to $\pi_{12,\gamma_{s}}$,   $ESS_{s+1}$ should not be that much different from $ESS_s$. On the other hand if that difference is pronounced then $ESS_{s+1}$ could drop dramatically. Hence in order to  determine the next auxiliary distribution, we define an  acceptable reduction in the $ESS$, i.e. $ESS_{s+1} \ge \zeta ~ESS_s$ (where $\zeta<1$) and prescribe $\gamma_{s+1}$ (\refeq{eq:m16}) accordingly.
The revised Adaptive SMC algorithm is summarized in Table \ref{tab:asmc}.

\begin{table}[htp!]
\begin{tabular}{|p{12cm}|}
\hline
\noindent {\bf Adaptive SMC algorithm:}
\begin{enumerate}
\item For $s=0$, let $\{\bs{\theta}_0^{(i)},~w_0^{(i)} \}_{i=1}^N$ be the initial  particulate approximation to $\pi_{12,\gamma_{0}}=\pi_1$ and $ESS_0$ the associated effective sample size. Set $s=1$.

\item {\em Reweigh}: If  $w_{s}^{(i)}(\gamma_s) =w_{s-1}^{(i)}~\frac{ \pi_{12,\gamma_{s}}(\bs{\theta}_{s-1}^{(i)}) }{ \pi_{12,\gamma_{s-1}}( \bs{\theta}_{s-1}^{(i)}) }$ are the {\em updated } weights as a function of $\gamma_s$ then determine $\gamma_s$ so that the associated $ESS_s = \zeta ESS_{s-1}$ (the value $\zeta=0.95$ was used in all the examples). Calculate $w_{s}^{(i)}$  for this $\gamma_s$.

\item{\em Resample}: If $ESS_s \le ESS_{min}$ then resample.

\item {\em Rejuvenate}: Use an MCMC kernel $P_s(.,.)$ that leaves $\pi_{12,\gamma_s}$
invariant to perturb each particle $\bs{\theta}_{s-1}^{(i)} \to \bs{\theta}_{s}^{(i)}$

\item The current population $\{\bs{\theta}_{s}^{(i)}, w_{s}^{(i)} \}_{i=1}^N$ provides a particulate approximation of $\pi_{12,\gamma_s}$  in the sense of Equations        (\ref{eq:m14}), (\ref{eq:m15}).

\item If  $\gamma_s < 1$ then set $s=s+1$ and goto to step 2. Otherwise stop.
\end{enumerate} \\
\hline
\end{tabular}
\caption{Basic steps of the {\em Adaptive } SMC algorithm proposed}
\label{tab:asmc}
\end{table}

\subsubsection{Trans-dimensional  MCMC}
\label{rjmcmc}

As mentioned earlier, a critical component in the SMC framework proposed is the MCMC-based rejuvenation step of the particles $\bs{\theta}$. It should be noted that the kernel $P_s(.,.)$ in the rejuvenation step (Step 3 of the SMC algorithm) need not be known explicitly as it does not enter in any of the pertinent equations. It is suffices that it is $\pi_{12,\gamma_s}$-invariant which is the target density. For the efficient exploration of the state space, we employ a mixture of moves which involve fixed dimension proposals (i.e. proposals for which the cardinality of the representation $k$ is unchanged) as well as moves which alter the dimension $k$ of the vector of parameters $\bs{\theta}$. We consider a total of $M=7$ such moves, each selected with a certain probability as discussed below. Of those, four involve trans-dimensional proposals which warrant a more detailed discussion. 

It is generally difficult to design proposals that alter the dimension significantly while ensuring a reasonable acceptance ratio. For that purpose, in this work we consider proposals that alter the cardinality  $k$ of the expansion by $1$ i.e. $k'=k-1$ or $k'=k+1$. We adopt the  the Reversible-Jump MCMC (RJMCMC) framework introduced in \cite{Green:1995} according to which such moves are defined in pairs in order to ensure reversibility of the Markov kernel (even though the reversibility  condition is not necessary, it greatly facilitates the formulations). We consider two such pairs of moves, namely {\em birth-death} and {\em split-merge}. Let  a proposal from $(k,\bs{\theta})$ to $(k',\bs{\theta'})$ that increases the dimension i.e.  $k'=k+1$ and  $\bs{\theta} \in \bs{\Theta}_k$, $\bs{\theta'} \in \bs{\Theta}^{k+1}$  (see last paragraph of sub-section \ref{prior}). Let $p(k \to k')$ the probability that such a proposal is made (user specified) and $p(k' \to k)$ the probability that the {\em reverse}, dimension-decreasing proposal is made. In order to account for the $m=dim(\bs{\Theta}_{k+1})-dim(\bs{\Theta}_k)$ difference  in the dimensions of  $\bs{\theta}$ and $\bs{\theta'}$, the former is augmented with a vector $\bs{u} \in \RR ^m$ drawn from a distribution $q(\bs{u})$. Consider a differential and one-to-one mapping $h: \bs{\Theta}_{k+1} \to \bs{\Theta}_{k+1} $ that connects the three vectors as $\bs{\theta'}=h(\bs{\theta},\bs{u})$. Then as it is shown in \cite{Green:1995}, the acceptance ratio of such a proposal is:
\be
\label{eq:rjmcmc}
\min \left\{ 1, \frac{\pi_{12,\gamma_s}(\bs{\theta'}) p(k \to k') }{ \pi_{12,\gamma_s}(\bs{\theta})p(k' \to k) }  \frac{1}{q(\bs{u})} \left| \frac{\partial \bs{\theta'}}{ \partial (\bs{\theta},\bs{u} )} \right| \right\}
\ee
where  $ \left| \frac{\partial \bs{\theta'}}{ \partial (\bs{\theta},\bs{u}) } \right| $ is the Jacobian of the mapping $h$. Such a proposal is invariant w.r.t. the density $\pi_{12,\gamma_s}$. Similarly one can define, the acceptance ratio of the {\em reverse}, dimension-decreasing move:
\be
\label{eq:rjmcmc1}
\min \left\{ 1, \frac{\pi_{12,\gamma_s}(\bs{\theta}) p(k' \to k) }{ \pi_{12,\gamma_s}(\bs{\theta'})p(k \to k') }  q(\bs{u}) \left| \frac{\partial \bs{\theta'}}{ \partial (\bs{\theta},\bs{u}) } \right|^{-1} \right\}
\ee

In the following we provide details for the reversible pairs used in this work.

\noindent {\bf Birth-Death:} In order to simplify the resulting expressions, we assign the following  probabilities of proposing one of these moves $p_{birth}=c~min\{1, \frac{p(k+1)}{p(k)} \}=c~ \frac{1}{s+1}$ (from \refeq{eq:m5a}) and $p_{death}=c~min\{1, \frac{p(k-1)}{p(k)} \}=c $ (from \refeq{eq:m5a}). The constant  $c$  is user-specified (it is taken equal to $0.2$ in this work). Obviously if $k=k_{max}$, $p_{birth}=0$ and if $k=0$, $p_{death}=0$.

For the death move:
\bi 
\item A kernel $j$ ($1 \le j \le k$ ) is selected uniformly and removed from the representation in \refeq{eq:m2}.
\item The corresponding $a_j$ is also removed.
\ei
For the birth move:
\bi
\item A new kernel $k+1$ is added to the expansion while the existing terms remain unaltered.
\item The associated amplitude $a_{k+1}$ is drawn from $\mathcal{N}(0,\sigma_4^2)$ (the variance $\sigma_4^2$ is equal to the average of the squared amplitudes $a_j$ over  all the particles at the previous iteration)
\item The associated scale parameter $\tau_{k+1}$ is drawn from the prior, \refeq{eq:m99}
\item The associated kernel location $\bs{x}_{k+1}$ is also drawn from the prior, \refeq{eq:m8}.
\ei
Hence the vector of dimension-matching parameters  $\bs{u}$ consists of $\bs{u}=(a_{k+1} ,$ $ \tau_{k+1}, \bs{x}_{k+1})$ and the corresponding proposal $q(\bs{u})$ is:
\be
\label{eq:rjmcmc2}
q(\bs{u})=\frac{1}{\sqrt{2\pi}} \frac{1}{\sigma_4}e^{-\frac{1}{2}~a_{k+1}^2/\sigma_4^2}~\frac{b_{\tau}^{a_{\tau}} }{\Gamma(a_{\tau}) } \tau_{k+1}^{ a_{\tau}-1 } ~\exp(-b_{\tau} \tau_{k+1})~\frac{1}{\mid \mathcal{D} \mid}
\ee
It is obvious that the Jacobian of such a transformation is $1$. 

\noindent {\bf Split-Merge} These moves correspond to splitting an existing kernel into two  or merging two existing kernels into one. Similarly to the birth-death pair, they alter the dimension of the expansion by $1$ and are selected with probabilities  $p_{split}= \frac{1}{s+1}$ and $p_{merge}=c $. For obvious reasons, $p_{split}=0$ if $k=k_{max}$ and $p_{merge}=0$ if $k \le 1$.    Consider first the  merge move between two kernels $j_1$ and $j_2$. 
 In order to ensure a reasonable acceptance ratio, merge moves are only permitted when the (normalized) distance between  the kernels is relatively small and when  the  amplitudes $a_{j_1}$, $a_{j_2}$  are relatively similar. Specifically we require that the following two conditions are met: 
\be
\label{eq:rjmcmc3}
\frac{\parallel \bs{x}_{j_1}-\bs{x}_{j_2} \parallel }{ \sqrt{ \tau_{j_1}^{-1}+\tau_{j_2}^{-1} } } \le \delta_x \qquad \mid a_{j_1}-a_{j_2} \mid \le \delta_a
\ee
(the values $\delta_x =\delta_a=1$ were used in this work). Two candidate kernels are selected {uniformly} from the pool of pairs satisfying the aforementioned conditions. The proposed kernels  $j_1$ and $j_2$ are removed from the expansion and are  substituted by a new kernel $j$ with the following associated parameters:
\bi
\item \be
\label{eq:rjmcmc4}
\tau_{j}= \left( \tau_{j_1}^{-1}+\tau_{j_2}^{-1}  \right)^{-1}
\ee
\item \be
\label{eq:rjmcmc5}
a_{j}=\sqrt{\tau_{j} }(\frac{a_{j_1}}{\sqrt{\tau_{j_1}} } + \frac{a_{j_2}}{\sqrt{\tau_{j_2}} } )
\ee
This ensures that the {\em average} value of the previous expansion (with $j_1$ and $j_2$) in \refeq{eq:m2} when integrated in $\RR^d$  is the same with the new (which contains $j$ in place of  $j_1$ and $j_2$)
\item \be
\label{eq:rjmcmc6}
\bs{x}_{j}=\frac{ \bs{x}_{j_1}+\bs{x}_{j_2} }{2}
\ee
\ei 

The split move is applied to a kernel $j$ (selected {\em uniformly}) which is substituted by two new kernels $j_1$, $j_2$.    In order to ensure {\em reversibility},  kernels $j_1$ and $j_2$ should satisfy the requirements of \refeq{eq:rjmcmc3} and the application of a merge move in the manner described above, should return to the original kernel $j$.  There are several ways to achieve this, corresponding essentially to different vectors $\bs{u}$ and mappings $h$ in \refeq{eq:rjmcmc}. In this work:
\bi
\item A scalar $u_{\tau}$ is drawn from the uniform distribution $U[0,1]$ and $\tau_{j_1}^{-1}=u_{\tau} \tau_j^{-1}$ and $\tau_{j_2}^{-1}=(1-u_{\tau}) \tau_j^{-1}$. This ensures compatibility with \refeq{eq:rjmcmc4}.
\item A vector $\bs{u}_x$ is drawn uniformly in the ball of radius $R$ where $R=\frac{\delta_x}{2\sqrt{\tau_j} }$. The center of the new kernels are specified as $\bs{x}_{j_1}=\bs{x}_j-\bs{u}_x$ and $\bs{x}_{j_2}=\bs{x}_j+\bs{u}_x$. This ensures compatibility with the first of \refeq{eq:rjmcmc3} as well as \refeq{eq:rjmcmc6}.
\item A scalar $u_a$ is drawn from the uniform distribution $U[-\frac{\delta_a}{2}, \frac{\delta_a}{2}]$. The amplitudes of the new kernels are determined by $a_{j_1}=\hat{a}-u_a$ and $a_{j_2}=\hat{a}+u_a$, where $\hat{a}=\frac{a+u_a(\sqrt{u_{\tau}}-\sqrt{1-u_{\tau} } ) }{ \sqrt{u_{\tau}}+\sqrt{1-u_{\tau} } }$. This ensures compatibility with the second of \refeq{eq:rjmcmc3} as well as \refeq{eq:rjmcmc5}.

\ei

The vector of dimension-matching parameters  $\bs{u}$ (in \refeq{eq:rjmcmc}) consists of $\bs{u}=(u_{\tau}, \bs{u}_x, u_a)$ and the corresponding proposal $q(\bs{u})$ is a product of uniforms in the domains specified above.
After some algebra, it can be shown that the Jacobian of such a transformation is $2^{d+1} \frac{\tau}{ u_{\tau}^2 (1-u_{\tau})^2 } \frac{1}{  \sqrt{u_{\tau}}+\sqrt{1-u_{\tau} } }$. 

\begin{figure}[]
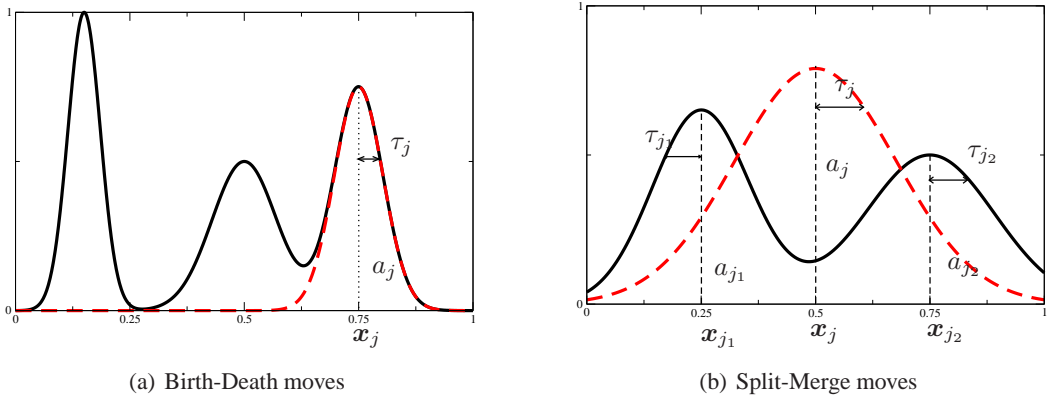

     \centering
     \subfigure[Birth-Death moves]{
          \label{fig:rjmcmc1}
\psfrag{a}{$a_j$}
\psfrag{t}{$\tau_j$}
\psfrag{x}{$\bs{x}_j$}
           \includegraphics[width=.45\textwidth]{FIGURES/rjmcmc1.eps}} \hfill
     \subfigure[Split-Merge moves]{
          \label{fig:rjmcmc2}
\psfrag{a}{$a_j$}
\psfrag{t}{$\tau_j$}
\psfrag{x}{$\bs{x}_j$}
\psfrag{a1}{$a_{j_1}$}
\psfrag{t1}{$\tau_{j_1}$}
\psfrag{x1}{$\bs{x}_{j_1}$}
\psfrag{a2}{$a_{j_2}$}
\psfrag{t2}{$\tau_{j_2}$}
\psfrag{x2}{$\bs{x}_{j_2}$}
             \includegraphics[width=.45\textwidth]{FIGURES/rjmcmc2.eps}}\\
     \label{fig:rjmcmc}
\caption{Trans-dimensional RJMCMC proposals}
\vspace{.5cm}
\end{figure}

The remaining three proposals, involve fixed-dimension moves that do not change the cardinality of the expansion but rather perturb  some of the terms involved. In particular, we considered updates of the amplitude $a_j$, scale $\tau_j$ or location $\bs{x}_j$ of a kernel $j$ selected {\em uniformly} (naturally, in the case of the amplitudes, the constant $a_0$ (\refeq{eq:m2}) is also a candidate for updating). Each of these three moves is proposed with probability $\frac{1}{3}(p_{birth}+p_{death}+p_{split}+p_{merge})=\frac{2~c}{3}(\frac{1}{s+1}+1)$. In particular:
\begin{enumerate} 
\item Update $a_j \to a'_j$: A coefficient $a_j$ (in \refeq{eq:m2}) is {\em uniformly} selected and perturbed as: 
\be
\label{eq:mcmc1}
a'_j =a_j + \sigma_1~Z \quad, Z\sim\mathcal{N}(0,1)
\ee
\item Update $\tau_j \to \tau'_j$: A scale parameter $\tau_j$ (in \refeq{eq:m2}) is {\em uniformly} selected and perturbed as: 
\be
\label{eq:mcmc2} 
\tau'_j =\tau_j e^{\sigma_2 Z}, \quad Z\sim\mathcal{N}(0,1)
\ee
(this ensures positivity of $\tau'_j$)
\item Update $\bs{x}_j \to \bs{x'}_j$: A location $\bs{x}_j \in \mathcal{D} \subset \RR^d$ (in \refeq{eq:m2}) is {\em uniformly} selected and perturbed as:
\be
\label{eq:mcmc3} 
\bs{x'}_j =\bs{x}_j+\sigma_3 ~\bs{Z}, \quad \bs{Z}=(Z_1,\ldots,Z_d), ~ Z_i \sim \mathcal{N}(0,1)
\ee
\end{enumerate}
The acceptance ratios are calculated based on the standard MCMC formulas using $\pi_{{12},\gamma_s}$ as the target density. 
It should be noted that the variances in the random walk proposals are adaptively selected so that the respective acceptance rates are in the range $0.2 - 0.4$. As it is well-known  (chapter 7.6.3 in \cite{rob04mon}) adaptive adjustments of Markov Chains based on  past samples can breakdown ergodic properties and lead to convergence issues in standard MCMC contexts. In the proposed SMC framework however, such restrictions do not apply as it suffices that the MCMC kernel is invariant. This is an additional advantage of the proposed simulation scheme in comparison to traditional MCMC.

\subsection{Prediction}
\label{prediction}

The significance of mathematical models for the computational simulation of  physical processes  lies in their {\em predictive ability}. It is these predictions that serve as the basis for engineering decisions in several systems of technological interest.
The proposed framework provides a seamless link from  experiments/data collection, to  model validation and ultimately prediction. In the presence of significant sources of uncertainty it is important not only to provide predictive estimates but quantify the level of confidence one can assign to the predicted outcome. 
 The inferred posteriors $\pi_r$ corresponding to various model resolutions can be used to carry out this task. In accordance with the Bayesian mind-set, all unknowns are considered random. If $\hat{\bs{y}}$  denotes the output to be predicted (under specified input, boundary \& initial conditions) then, the {\em predictive posterior}  $p(\hat{\bs{y}} \mid \bs{y})$ based on the available data $\bs{y}$ can be expressed as (\cite{gel03bay}):
\begin{eqnarray}
\label{eq:m01}
p(\hat{\bs{y}} \mid \bs{y}) & = & \int p(\hat{\bs{y}}, \bs{\theta} \mid \bs{y}) ~d\bs{\theta} =  \int  p_r(\hat{\bs{y}} \mid  \bs{\theta}, \bs{y}) \underbrace{ p( \bs{\theta} \mid \bs{y})}_{\textrm{posterior}}  ~d\bs{\theta} \\ \nonumber
& = &  \int  \underbrace{ L_r(\hat{\bs{y}} \mid  \bs{\theta} )}_{\textrm{likelihood}} \pi_r ( \bs{\theta} ) ~d\bs{\theta} \approx \sum_{i=1}^N W_r^{(i)}~L_r(\hat{\bs{y}} \mid \bs{\theta}_r^{(i)} )
\end{eqnarray}
The term $p(\hat{\bs{y}} \mid  \bs{\theta} )$ is the likelihood of the predicted data determined by the forward solver at resolution $r$ as in \refeq{eq:m9c}. \refeq{eq:m01} offers an intuitive interpretation of the predictive process. The predictive posterior distribution is a mixture of the corresponding likelihoods evaluated at all possible  states $\bs{\theta}$ of the system , with weights proportional to the their posterior values. In the context of Monte Carlo simulations, samples of $\hat{\bs{y}}$ from  $p(\hat{\bs{y}} \mid \bs{y}) $ can be readily drawn using the  particulate approximation of each $\pi_r$ (\refeq{eq:m14}). These samples can subsequently be used to  statistics of the predicted output $\hat{\bs{y}}$  such as moments, probabilities of exceedance which can be extremely useful in engineering practice.

%% file: results_short.tex
The method proposed is illustrated in problems  from nonlinear  solid mechanics using artificial data. The governing PDEs are those of small-strain, rate-independent, perfect plasticity with a von-Mises yield criterion and associative flow rule (\cite{sim00com}):
\bea
\nabla \cdot  \bs{\sigma}(\bs{x})  =  0  &   & \textrm{ \hfill (conservation of
linear momentum)} \nonumber \\
\bs{\sigma}  = \bs{C}(E,\nu) : (\bs{\epsilon} -\bs{\epsilon}^{p} ) & &  \textrm{
 \hfill (elastic stress-strain relationships)} \nonumber \\
h(\bs{\sigma}):  =  \sqrt{ \parallel \bs{\sigma} \parallel^2-\frac{1}{3}(tr[\bs{
\sigma}])^2 } -\sqrt{\frac{2}{3}} \sigma_{yield}  & &  \textrm{ \hfill (yield surface)} \nonumber \\
\bs{\dot{\epsilon}}^p  = \lambda \frac{\partial h}{\partial \bs{\sigma}}  & &  \textrm{\hfill (flow~ rule)} \nonumber \\
\label{eq:plast}
\eea
where $\bs{\sigma}$ is the Cauchy stress-tensor, $\bs{\epsilon}=\frac{1}{2}(\nabla u+u \nabla) $ and $\bs{\epsilon}^{p}$ the total and plastic-part of the strain tensor, $\bs{v}=(v_x,v_y,v_z)$ is the displacement vector, $\bs{C}(E,\nu)$ is the elastic moduli which depends on the Young's modulus $E$ (it was assumed that it was known $E=1,000$) and Poisson's ratio $\nu$ (it was assumed that it was
 known $v=0.3$). The field of interest in all the  problems examined  was the yield stress $\sigma_{yield}(\bs{x})$ which was assumed to vary spatially. The yield stress determines the boundary of the elastic domain in the material response.  
A square two-dimensional domain $\mathcal{D}=[0,1]\times [0,1]$ under plane stress conditions was considered and the forward solvers were  Finite Element models which discretize the governing PDEs of \refeq{eq:plast} for $\bs{x} \in \mathcal{D}$. In order to construct a sequence of solvers operating at different resolutions, we considered $4$  different partitions corresponding to uniform $8 \times 8$, $16 \times 16$, $32 \times 32$ and $64 \times 64$ grids (i.e. with element sizes $\frac{1}{8} \times \frac{1}{8}$, $\frac{1}{16} \times \frac{1}{16}$, $\frac{1}{32} \times \frac{1}{32}$ and $\frac{1}{64} \times \frac{1}{64}$ respectively).   A critical issue with   spatially varying parameters is how this variability is accounted in the  discretized representation. In this work, we adopted a simple rule according to which each finite element was assigned a constant yield stress value  which was equal to the average of the field  $\sigma_{yield}(\bs{x})$  within the element. This scheme by no means represents a consistent upscaling of the governing PDEs let alone being optimal. It can be easily established that it can introduce significant deviations in the effective response which depends on the full details of the spatially varying field. 
 This poor selection is made however to emphasize  the point that inaccurate solvers can  be useful and can lead to significant improvements in accuracy and efficiency.  Their role is to provide a computationally inexpensive approximation to the fine-scale posterior that can be efficiently updated and refined using a reduced number of runs from more expensive solvers. Naturally, if more sophisticated upscaling schemes are introduced, the transitions in the sequence of posterior become smoother and the computational effort is further reduced.
Since $\sigma_{yield}(\bs{x})> 0 ~\forall \bs{x}$, we used our model  to  infer $log (\sigma(\bs{x}))$ i.e. in \refeq{eq:m2}, $f(\bs{x})=log(\sigma(\bs{x}))$.
The adaptive SMC scheme (Table \ref{tab:asmc})  with  $N=100$ or $N=500$ particles was employed in  the examples presented with $\zeta=0.95$ and $ESS_{min}=N/2$. 
The following values for the hyperparameters of the prior model were used (section \ref{prior}):
\bi
\item $k_{max}=100$ and $s=0.1$ (\refeq{eq:m5a})
\item $a_{\tau}=1.0$ (\refeq{eq:m9}) and $a_{\mu}=0.0001$ (\refeq{eq:m99})
\item $a_0=1.0$ and $b_0=1.0$ (\refeq{eq:m6a})
\item $a=2.$ and $b=1. \times 10^{-6}$ (\refeq{eq:m9c})
\ei

\subsection{Example A}
\begin{figure}
     \centering
     \subfigure[2D view]{
          \label{fig:ex31a}
           \includegraphics[width=.48\textwidth,height=6cm]{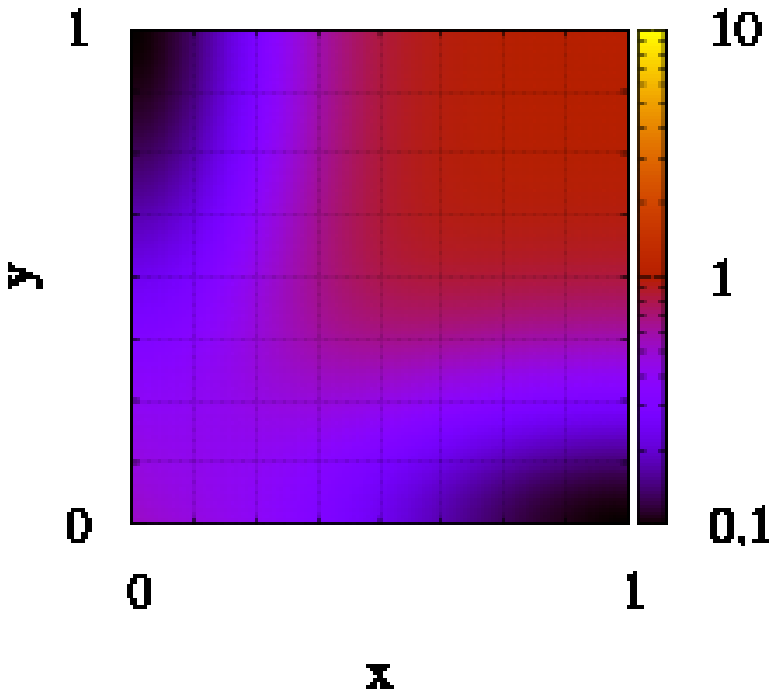}} \hfill
     \subfigure[3D view]{
          \label{fig:ex31b}
             \includegraphics[width=.48\textwidth,height=6cm]{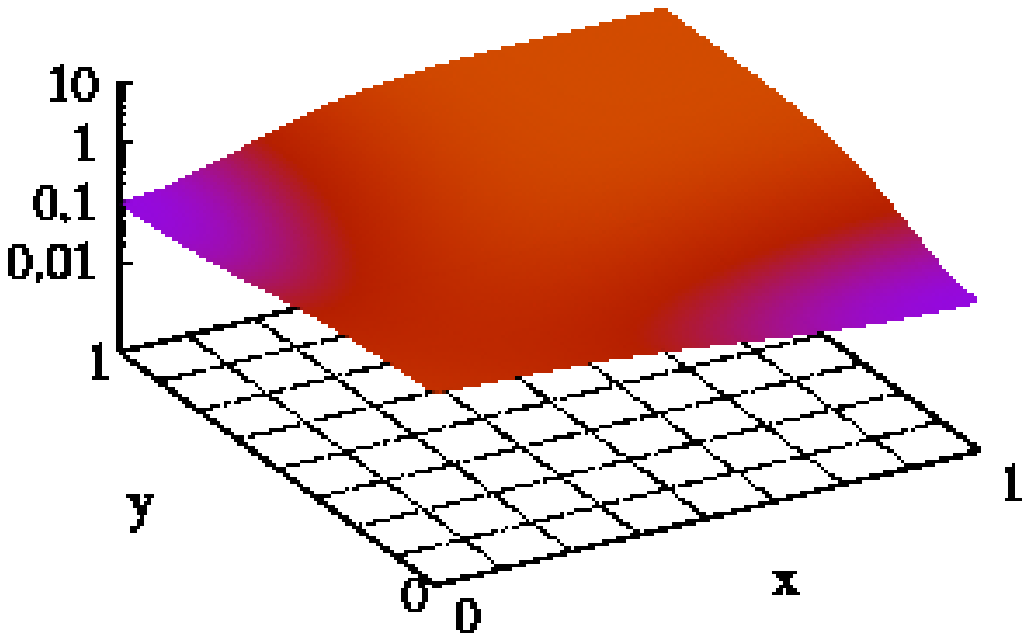}}\\
     \caption{Reference $\sigma_{yield}(\bs{x})$ field for Example A}
     \label{fig:ex31}
\end{figure}

In this example it was assumed that the yield stress varied as follows (Figure \ref{fig:ex31}):
\be
\label{eq:ex31}
\log \sigma_{yield}(\bs{x})=-1~\exp\{-10~x^2-2~(y-1)^2\}-1~\exp\{-2~(x-1)^2-10~y^2\}
\ee

The nonlinear governing PDEs (\refeq{eq:plast}) were solved using a $64 \times 64$ uniform finite element mesh with the following boundary conditions:
\bi
\label{eq:ex32}
\item $v_x=v_y=0$ along $x=0$
\item $v_x=-v_y=0.001$ along $x=1$
\ei
The displacements $v_x$, $v_y$ at a regular grid consisting of $72$  points with coordinates $(0.125~i,~ 0.125~j)$, for  $i=1,\ldots,8$ and $j=0,\ldots,8$ were recorded resulting in $n=144$ data points (as in Figure \ref{fig:ex31}).
The empirical mean  (of the absolute values) of these  observations $\mu_A$ was calculated and the recorded values were contaminated by Gaussian noise of standard deviation $5\%~\mu_A$ in order to obtain  sets of  {\em observables} denoted by $\{ y_i\}_{i=1}^n$ in our Bayesian model (\refeq{eq:m1}).
We note that in this example the scale of variability  of the unknown field $\sigma_{yield}(\bs{x})$ is {\em larger} than the scale of observations, i.e. the grid size where displacements were recorded.

Table \ref{tab:ex31} reports the number of degrees of freedom per solver and the normalized computational time for a single run w.r.t. the $64 \times 64$ solver.
As mentioned earlier,  each finite element was assigned a constant yield stress equal to the average value inside the  element. This is of course inconsistent with the governing PDEs as the geometry of the variability plays a critical role for the effective properties of each element. It is easily understood though that the corresponding posterior should have some similarities arising from the mere nature of their construction.

\begin{table}
\begin{tabular}{|c|c|c|}
\hline \hline
Solver & Degrees of & Normalized  Computational \\
Resolution &  Freedom & Time (Actual in sec) \\
\hline
$16 \times 16$ & $510$ &$\frac{1}{156}$ (0.55) \\ \hline
$32 \times 32$ & $2,046$  &$\frac{1}{18}$ (4.8)  \\ \hline
$64 \times 64$ & $8,190$ &$1$ (86)  \\
\hline \hline
\end{tabular}
\caption{Computational cost of different resolution solvers for Example A }
\label{tab:ex31}
\end{table}

\begin{figure}
     \centering
     \subfigure[Posterior mean]{
          \label{fig:ex32a}
           \includegraphics[width=.48\textwidth,height=6cm]{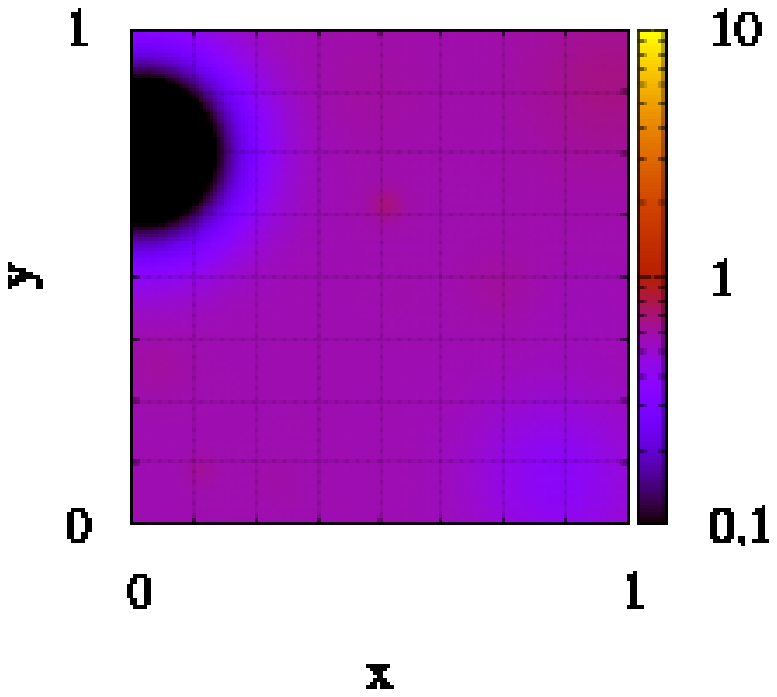}} \hfill
     \subfigure[Posterior $5\%$ and $95\%$ quantiles]{
          \label{fig:ex32b}
             \includegraphics[width=.48\textwidth,height=6cm]{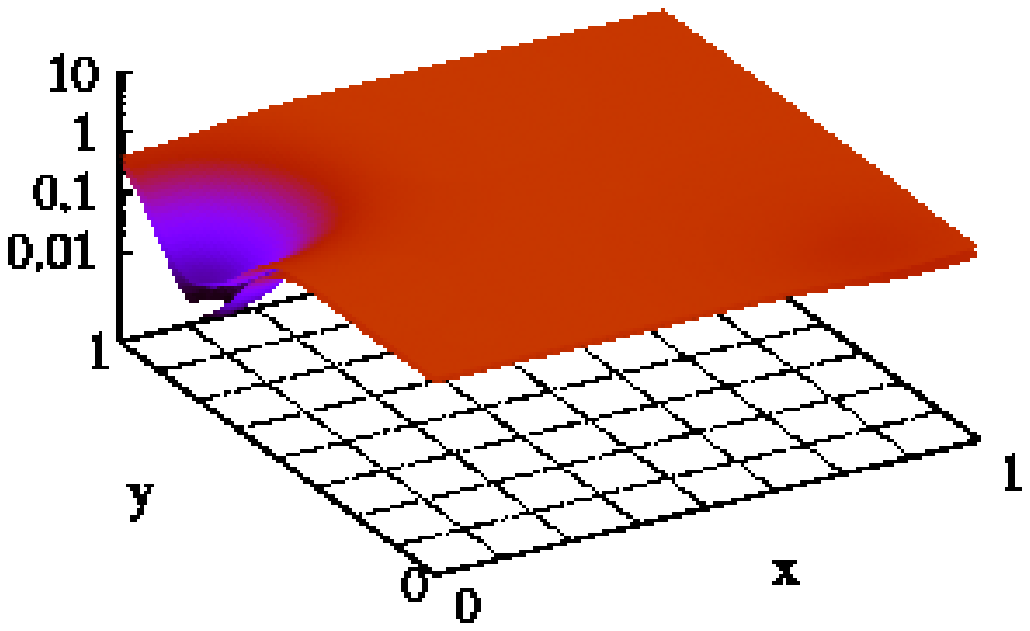}}\\
     \caption{Posterior inference using only the $64 \times 64$ solver}
     \label{fig:ex32}
\end{figure}

At first, we attempted to solve the problem by  operating solely on the finest solver. Using the Adaptive SMC scheme proposed with $N=100$ particles, this resulted in a sequence of $163$ (between the prior $\pi_0$ and the target posterior) auxiliary bridging distributions constructed as mentioned earlier. The inferred field (posterior mean and quantiles) are depicted in Figure \ref{fig:ex32}. Even though they exhibit similarities with the ground truth (Figure \ref{fig:ex31}), there are also considerable differences  which suggest that the algorithm probably   got trapped in some mode of the posterior. 
This is to be expected due to the highly nonlinear nature of the forward solver and the large state space. It is possible however that the correct solution could be recovered if the size of the population and/or the number of bridging distributions is increased. Inspite of that, it is the significant  computational effort  that makes such an approach impractical. In particular $16,300$ (i.e. $163 \times 100$) calls to the most expensive forward solver were required.

\begin{figure}
     \centering
     \subfigure[Resolution $16\times 16$ - quantile $5\%$]{
          \label{fig:ex33a}
           \includegraphics[width=.48\textwidth,height=6.cm]{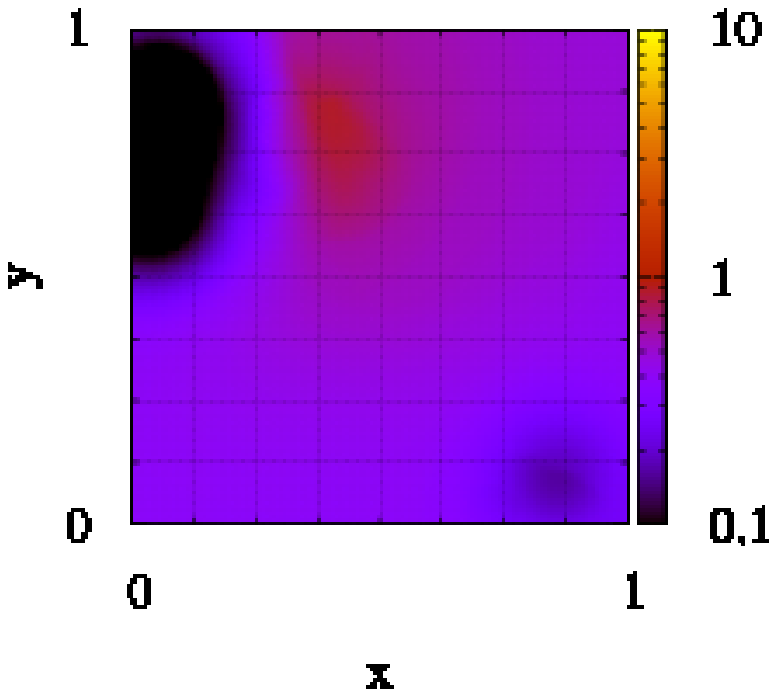}} \hfill
     \subfigure[Resolution $16\times 16$ - quantile $95\%$]{
          \label{fig:ex33b}
             \includegraphics[width=.48\textwidth,height=6.cm]{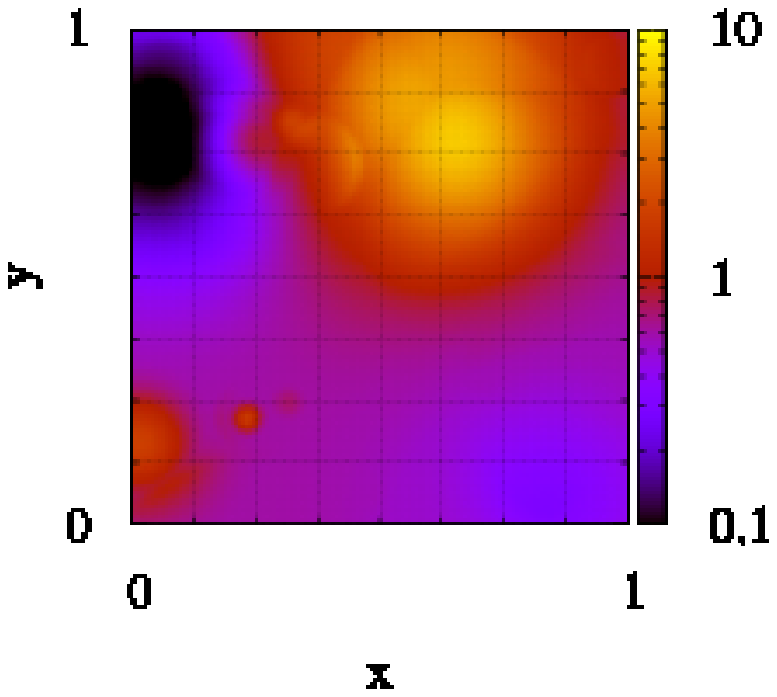}}\\
     \subfigure[Resolution $32\times 32$ - quantile $5\%$]{
          \label{fig:ex33c}
           \includegraphics[width=.48\textwidth,height=6.cm]{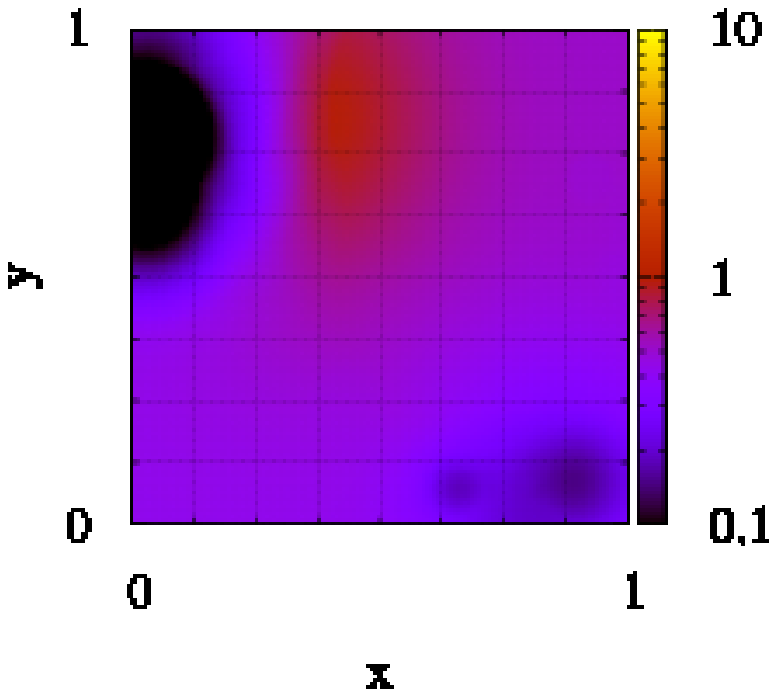}} \hfill
     \subfigure[Resolution $32\times 32$ - quantile $95\%$]{
          \label{fig:ex33d}
             \includegraphics[width=.48\textwidth,height=6.cm]{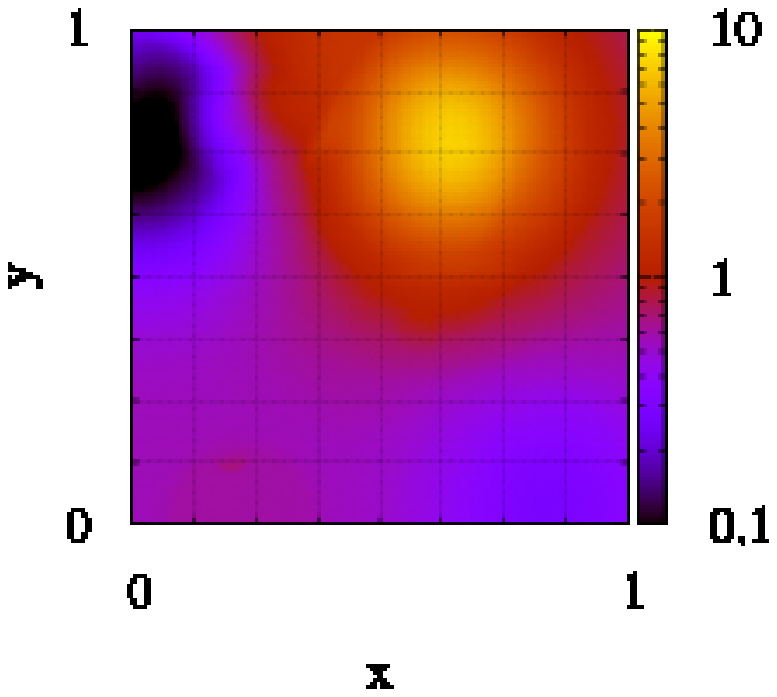}}\\
     \subfigure[Resolution $64\times 64$ - quantile $5\%$]{
          \label{fig:ex33e}
           \includegraphics[width=.48\textwidth,height=6.cm]{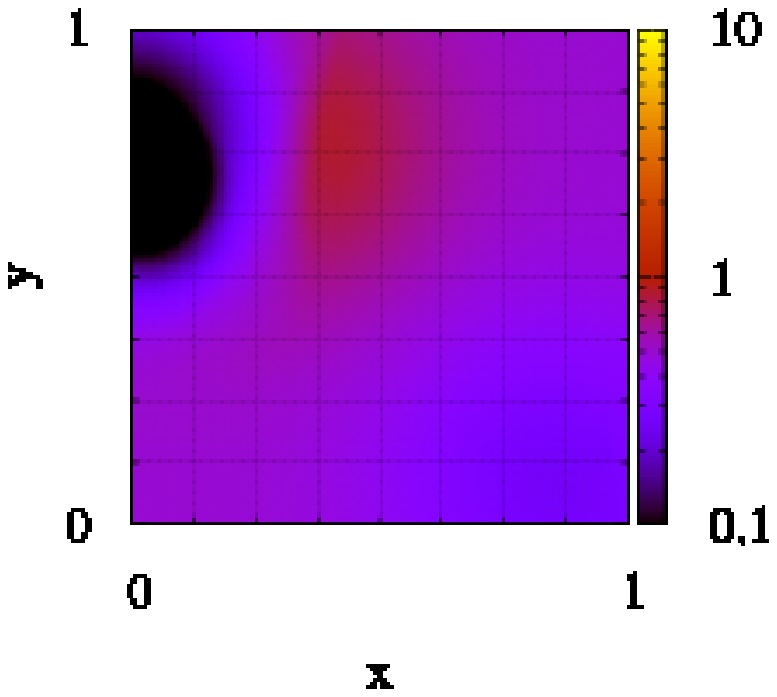}} \hfill
     \subfigure[Resolution $64\times 64$ - quantile $95\%$]{
          \label{fig:ex33f}
             \includegraphics[width=.48\textwidth,height=6.cm]{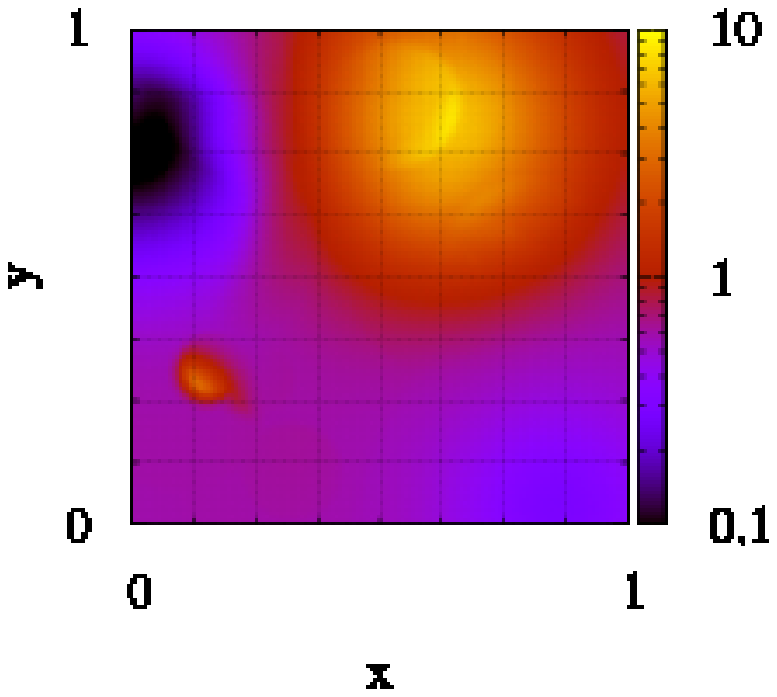}}\\
     \caption{Posterior quantiles at various solver resolutions for Example A}
     \label{fig:ex33}
\end{figure}

In contrast, when a sequence of $3$ solvers was used the results obtained are significantly closer to the ground truth as it can be seen in Figures \ref{fig:ex33} and  7. 
It is observed that even using the coarsest solver ($16 \times 16$), we are able to correctly identify some of the basic features of the underlying field. The inferences are greatly improved as solvers at finer resolutions are invoked. Figure 8 
 depicts the number of bridging distributions needed at each resolution and the respective reciprocal temperatures $\gamma_s$ (\refeq{eq:m16}). These were {\em automatically determined } by the proposed Adaptive SMC with $N=100$ particles. It is also observed  that the number of intermediate distributions needed decreased as finer resolution solvers are used. This is a direct consequence of the ability of the proposed scheme to accumulate  information from coarser scale solver. These results are  summarized in Table \ref{tab:ex32} which also reports the {\em effective} computational cost at the various stages and in total. It can be seen that a reduction of the total number of calls is achieved ($16,300$ vs. 6,265$ $).

Figure \ref{fig:ex36} depicts the posterior densities of the inferred model error standard deviations $\sigma_r$ described in \refeq{eq:m9b}. It is readily seen that the proposed technique is  able to quantify the magnitude of the model error for solvers of various resolutions. Furthermore for the reference resolution $64\times 64$ it correctly detects that the error contamination is of the level of $5\% \mu_A$.
Finally Figure \ref{fig:ex37} depicts the marginal posterior on $k$ , i.e. the cardinality of the expansion at various resolutions. It should be noted that the method leads to {\em sparse} representations (on average $k=5$ and therefore only $21$ parameters are needed) without sacrificing the accuracy. Traditional formulations (deterministic or probabilistic) usually  have as many unknowns as elements (i.e. in the $64 \times 64$ mesh, $4,096$ parameters) and therefore  require operations in very high dimensional spaces with all the negative implications this carries.

\begin{figure}
     \centering
    \subfigure[Resolution $16\times 16$]{
          \label{fig:ex35a}
             \includegraphics[width=.48\textwidth,height=6cm]{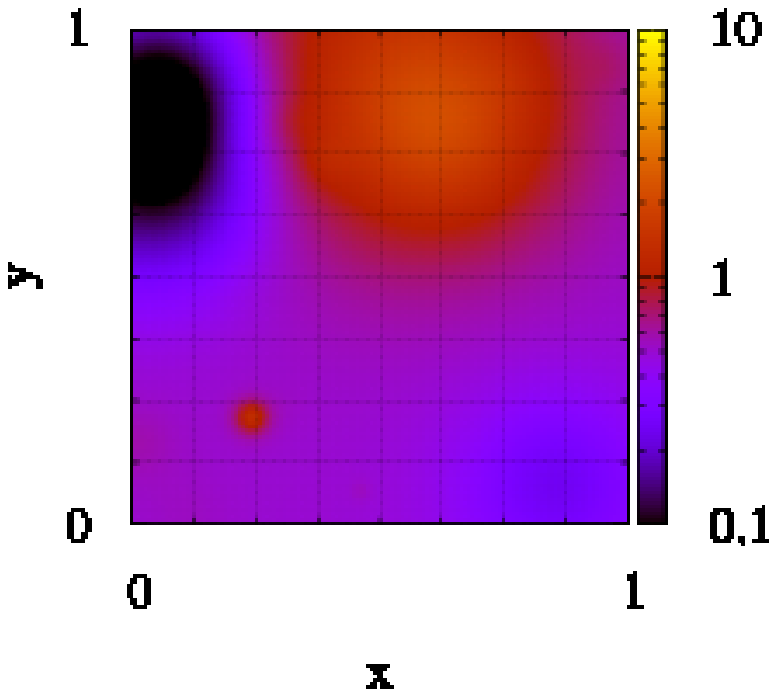}}\hfill
     \subfigure[Resolution $32\times 32$]{
          \label{fig:ex35b}
           \includegraphics[width=.48\textwidth,height=6cm]{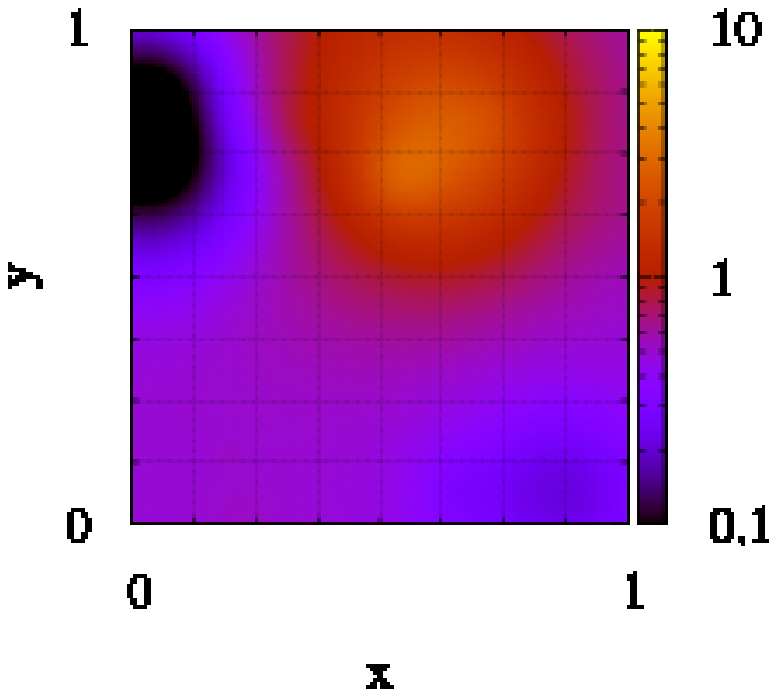}} \\
     \subfigure[Resolution $64\times 64$ ]{
          \label{fig:ex35c}
             \includegraphics[width=.48\textwidth,height=6cm]{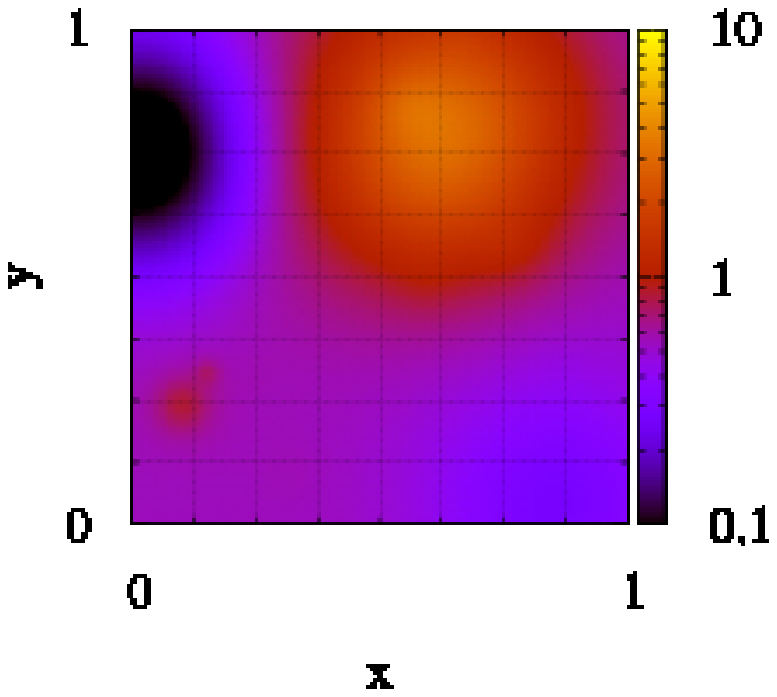}}\\

     \caption{Posterior mean at various solver resolutions for Example A}
 \label{fig:ex35}
\vspace{0.5cm}
\end{figure}

\begin{figure}
\label{fig:ex34}
\centering
\psfrag{gamma}{$\gamma_s$}
\psfrag{iterations}{iterations}
\includegraphics[width=.8\textwidth,height=6cm]{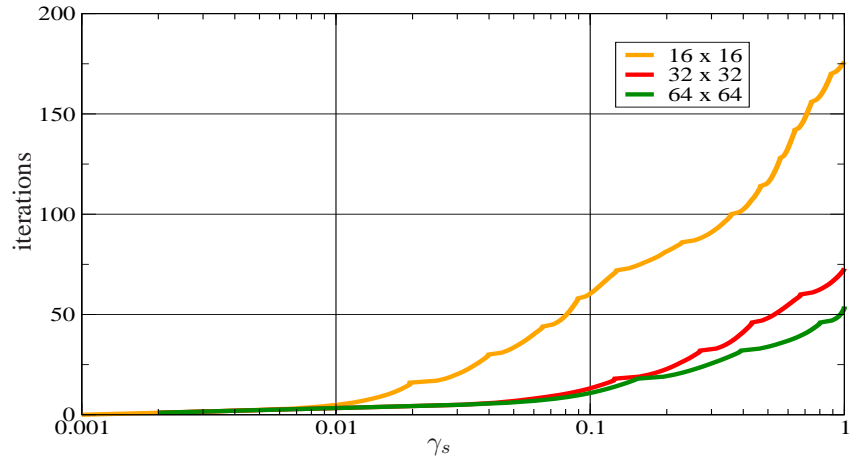}
\caption{Evolution of reciprocal temperature $\gamma_s$  (\refeq{eq:m16}) and number of bridging distributions}
\vspace{.5cm}
\end{figure}

\begin{table}
\begin{tabular}{|c|c|c|}
\hline \hline
Solver & Number of Bridging  & Computational Effort \\
Resolution &  Distributions & (w.r.t. calls to $64\times64$ solver) \\
\hline
$16 \times 16$ & $176$ & $113$ \\ \hline
$32 \times 32$ & $73$ & $452$   \\ \hline
$64 \times 64$ & $54$ & $5,700$   \\
\hline \hline
Total &   &  $6,265$ \\
\hline \hline    
\end{tabular}
\caption{Computational cost for inferences for Example A. Note that the effective cost when using only the $64\times 64$ solver was $16,300$ }
\label{tab:ex32}
\end{table}

\begin{figure}
\psfrag{r16}{\tiny $r=16\times 16$}
\psfrag{r32}{\tiny $r=32\times 32$}
\psfrag{r64}{\tiny $r=64\times 64$}
\includegraphics[width=.8\textwidth,height=5cm]{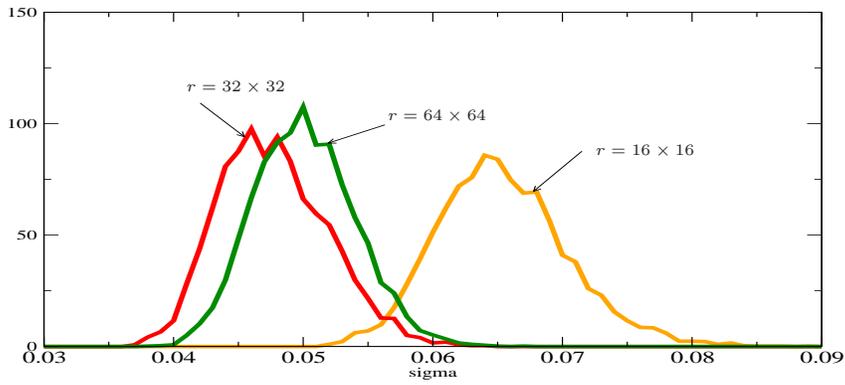}
\caption{Posterior densities of model error st. deviations $\sigma_r$ as in \refeq{eq:m9b}. The values on $x$-axis have been divided by  $\mu_A$ }
\label{fig:ex36}
\vspace{.5cm}
\end{figure}

\begin{figure}
\psfrag{k}{$k$}
\psfrag{ir16}{$16 \times 16$}
\psfrag{ir32}{$32 \times 32$}
\psfrag{ir64}{$64 \times 64$}
 \includegraphics[width=.8\textwidth,height=5.5cm]{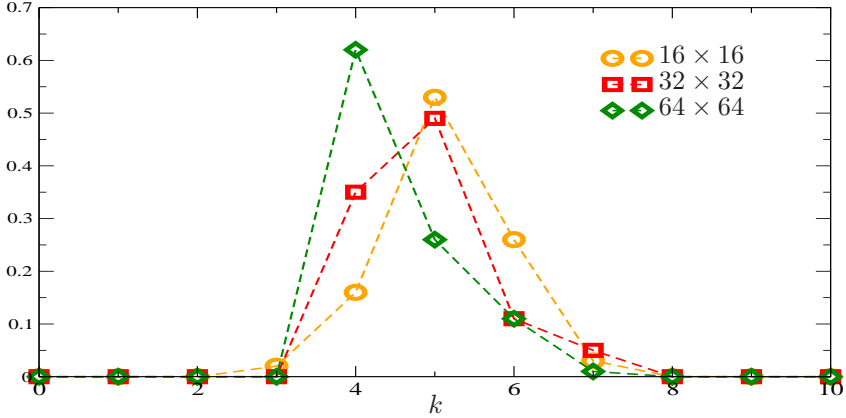}
     \caption{Posterior for the cardinality $k$ of the field representation}
     \label{fig:ex37}
\vspace{.5cm}
\end{figure}

%% file: conclusions.tex
A general Bayesian framework has been presented for the identification of spatially varying model parameters. The proposed model utilizes a parsimonious, non-parametric formulation that favors sparse representations and whose complexity can be determined from the data. An efficient inference scheme based on SMC has been discussed which is embarrassingly parallelizable and well-suited for detecting multi-modal posterior distributions. They key element is the introduction of an appropriate sequence of posteriors based on a natural hierarchy introduced by various forward solver resolutions. As a result, inexpensive, coarse solvers are used to identify the most salient features of the unknown field(s) which are subsequently enriched by invoking solvers operating at finer resolutions. The overall computational cost is further reduced by employing a novel adaptive scheme that automatically determines the number of intermediate steps. 
The proposed methodology does not require that Markov Chains using all the  solvers to be run simultaneously as in other multi-resolution formulations (\cite{hig03mar}) . The particulate approximations provide a concise way of representing the posterior which can be readily updated if the analyst wants to employ forward models operating at even finer resolutions or in general more accurate  solvers. The output of the inference algorithm provides estimates of the model error or noise contained in the data. An important feature is the ability to readily provide  not only predictive estimates but also quantitative measures of the {\em predictive uncertainty}. Hence it offers a  seamless link between data, computational models and predictions. 
The efficiency of the sampling schemes proposed could be greatly improved if the proposed moves incorporate information about the governing PDEs and if upscaling relations are available. 
A feature that was not explored in the examples presented is the possibility of performing {\em adaptive refinement}, not for the purposes of improving the forward solver accuracy but rather for increasing the resolution of  the unknown fields. This can be achieved in two ways and is a direct consequence of the ability of the proposed model (and Bayuesian models in general) to  produce  credible intervals for the estimates made at each step. Hence in regions where the variance of the estimates (or some other measure of random variability) is high, the resolution of the forward solver can be increased. Furthermore, additional measurements/data can be obtained at these regions if such a possibility exists. Hence the proposed framework allows for  near-optimal use of the computational resources and sensors  available.